\documentclass[aps,pre,reprint,superscriptaddress,twocolumn,longbibliography,floatfix]{revtex4-2}

\usepackage{graphicx}
\usepackage{dcolumn}
\usepackage{bm}

\usepackage[T1]{fontenc}
\usepackage{etoolbox}
\usepackage{xcolor}
\usepackage{mathpazo}

\usepackage[unicode=true,
bookmarks=true,bookmarksnumbered=false,bookmarksopen=false,
breaklinks=false,pdfborder={0 0 0},pdfborderstyle={},backref=false,colorlinks=true]
{hyperref}

\hypersetup{pdfborderstyle={},pdfborderstyle={},pdfborderstyle={},pdfborderstyle={},pdfborderstyle={},pdfborderstyle={},linkcolor=blue,citecolor=blue}
\usepackage{amssymb,amsmath}
\usepackage{comment}

\usepackage{float}

\newcommand{\oL}{\overline{L}}
\newcommand{\oX}{\overline{X}}
\newcommand{\oxi}{\overline{\xi}}
\newcommand{\oXi}{\overline{\Xi}}
\newcommand{\oR}{\overline{R}}
\newcommand\beq{\begin{equation}}
\newcommand\eeq{\end{equation}}
\newcommand\beqa{\begin{eqnarray}}
\newcommand\eeqa{\end{eqnarray}}
\newcommand{\nn}{\nonumber\\}
\def\bal#1\eal{\begin{align}#1\end{align}}
\newcommand{\hs}{d}
\newcommand{\shs}{\text{SHS}}

\newcommand{\ones}{\text{\textsf{\textbf{1}}}}

\newcommand{\figureheight}{1.4\columnwidth}

\newcommand{\SOne}{\ref{fig12_Alex}}\newcommand{\STwo}{\ref{Delta0p1_rho0p1}}\newcommand{\SThree}{\ref{Delta0p1_rho0p2}}\newcommand{\SFour}{\ref{Delta0p1_rho0p3}}\newcommand{\SFive}{\ref{Delta0p1_rho0p4}}\newcommand{\SSix}{\ref{htilde_Delta0p1_rho0p1}}\newcommand{\SNine}{\ref{htilde_Delta0p1_rho0p4}}\newcommand{\STen}{\ref{fig1aT2_Munao}}\newcommand{\SEleven}{\ref{fig1aT1p5_Munao}}\newcommand{\STwelve}{\ref{fig1bT2_Munao}}\newcommand{\SThirteen}{\ref{fig1bT1p5_Munao}}\newcommand{\SFourteen}{\ref{SS12_Delta0p25_rho0p3}}\newcommand{\SFifteen}{\ref{SS11_Delta0p5_rho0p3}}\newcommand{\SSixteen}{\ref{fig6_Riccardo}}\newcommand{\SNineteen}{\ref{fig10_Riccardo}}

\makeatletter
\def\@email#1#2{%
 \endgroup
 \patchcmd{\titleblock@produce}
  {\frontmatter@RRAPformat}
  {\frontmatter@RRAPformat{\produce@RRAP{*#1\href{mailto:#2}{#2}}}\frontmatter@RRAPformat}
  {}{}
}%
\makeatother

\newcommand{\figheight}{1.4\linewidth}
\newcommand{\largefigheight}{2.1\linewidth}
\newcommand{\mpwidth}{0.49\columnwidth}
\makeatletter
\providecommand{\captionof}[1]{\def\@captype{#1}\caption}
\makeatother

\begin{document}

\title{A Unified Rational-Function Approximation for Square-Well, Square-Shoulder, and Nonadditive Hard-Sphere Mixtures}
\author{Andr\'es Santos}%
\author{Santos B. Yuste}
 \affiliation{Departamento de F\'isica, Universidad de Extremadura, E-06006 Badajoz, Spain}
 \affiliation{Instituto de Computaci\'on Cient\'ifica Avanzada (ICCAEx), Universidad de Extremadura, E-06006 Badajoz, Spain}
\author{Ana M. Montero}
 \affiliation{Departamento de F\'isica, Universidad de Extremadura, E-06006 Badajoz, Spain}
\author{Mariano L\'opez de Haro}
\affiliation{Instituto de Energ\'{\i }as Renovables, U.N.A.M.,
Privada Xochicalco s/n, Col. Centro, Temixco, Mor.\ 62580, Mexico}

\date{\today}

\begin{abstract}
We develop a rational-function approximation for the structural properties of multicomponent fluids with narrow square-well and square-shoulder interactions, together with weakly nonadditive hard-sphere mixtures. The theory is obtained by reformulating the analytical Percus--Yevick solution for additive sticky hard-sphere mixtures so as to account for finite interaction widths, leading to a unified semianalytical framework for these classes of fluids. The approximation satisfies the exact low-density limit and preserves the continuity of the cavity functions; a simple local correction restores the continuity of their first derivatives. Comparison with Monte Carlo simulations for binary mixtures shows that the theory accurately describes both the radial distribution functions and their Fourier transforms for narrow wells and shoulders and for weak nonadditivity. The proposed formulation extends previous sticky-hard-sphere-based approaches while retaining their analytical simplicity.

\end{abstract}

\maketitle

\section{Introduction}

Relatively simple models of intermolecular interactions have long served as valuable tools for investigating the thermodynamic and structural properties of fluids within statistical mechanics.
Prominent among such models  are those with a hard-core, namely the hard-sphere (HS) potential, the square-well (SW) potential, and the square-shoulder (SS) potential, which capture the essential features of many real systems.
There is already a vast amount of literature in this field devoted mainly to monocomponent fluids,
although fluid mixtures have been investigated in parallel since the early stages \cite{H74,LSP86,LC87,LDS89,HF89,WW90,GWL90,H92b,GJMR94,RGM95,BD95,CW00,CW01,VBG01,WL07b,DFV07,DFMV09} and interest in them has increased in recent times \cite{DFYV14,F16b,MRB16,AS19,MTHG22,PPCMM23,MCMBP23}.
Part of this interest stems from the fact that some soft-matter systems, such as colloids and liquid metals, may be modeled as  SW or  SS mixtures.

In previous work, we  have used the so-called rational-function approximation (RFA) formalism \cite{S16} to obtain the structural properties of HS fluids \cite{YS91,YHS96,HSY06,HYS08,SYH20}, sticky hard-sphere (SHS) fluids \cite{YS93a,YS93b}, SW fluids \cite{YS94,AS01,LSAS03,LSYS05}, SS fluids \cite{YSH11,HYS16}, fluids interacting via piecewise constant potentials \cite{SYH12,SYHBO13,HRYS18,YSH22,MYSH25}, additive hard-sphere (AHS) fluid mixtures \cite{YSH98,YSH00a,MMYSH02,PBYSH20,PYSHB21,PYSHB21b},  and SHS mixtures \cite{SYH98}. A closely related RFA was previously proposed for nonadditive hard-sphere (NAHS) mixtures \cite{FS11,FS13}.

The aim of this paper is to formulate a theoretical framework within the RFA approach capable of describing the structural properties  of narrow SW, SS, and even NAHS mixtures.
The central idea is to reinterpret the analytical Percus--Yevick (PY) solution for SHS mixtures \cite{PS75,B75,BT79} as the starting point for constructing an RFA for finite-width interaction shells.
An attractive feature of the resulting formulation is that the same analytical framework applies, with only minor changes in the interaction parameters, to three different classes of fluid mixtures.

While, by construction, the applicability of our nonperturbative semi-analytical method is limited to narrow SW/SS coronas or to weak nonadditivities, it still provides a qualitatively reasonable description outside its nominal range of applicability.

The structure of the paper is the following. In Sec.~\ref{sec2}, we introduce the intermolecular potential and the main structural properties associated with this potential. This is followed in Sec.~\ref{sec3} by the explicit consideration of the SHS mixture in the PY approximation. Section~\ref{sec4} presents our proposal for the RFA of narrow SW/SS mixtures and NAHS mixtures. Next, Sec.~\ref{sec5} is devoted to a comparison between theory and simulation data for a variety of systems. Finally, Sec.~\ref{sec6} contains some concluding remarks.
Mathematical details of some derivations are provided in the Appendix, while additional comparisons with simulation data, including a ternary mixture, are presented in the Supplemental Material (SM)~\cite{SM_26}.

\section{Intermolecular potential and basic structural properties}
\label{sec2}

Let us consider an $n$-component mixture of spherical particles with a number density $\rho_i$ of component $i=1,\ldots,n$. The particles of species $i$ have a hard-core diameter $\sigma_i$ and an interaction range $\lambda_i$. The mole fraction of species $i$ is $x_i=\rho_i/\rho$, where $\rho=\sum_i\rho_i$ is the total number density. Their interaction potential is given by
\beq
\label{1}
\phi_{ij}(r)=\begin{cases}
\infty,&r<\sigma_{ij},\\
\epsilon_{ij},&\sigma_{ij}<r<\lambda_{ij},\\
0,&r>\lambda_{ij}.
\end{cases}
\eeq
Here, $r$ is the center-to-center distance and $\sigma_{ij}=\frac{1}{2}(\sigma_i+\sigma_j)$ is the additive contact distance between particles of species $i$ and $j$. We further assume that $\lambda_{ij}=\frac{1}{2}(\lambda_i+\lambda_j)$, so that the interaction-shell widths
\beq
\Delta_{ij}\equiv\lambda_{ij}-\sigma_{ij}
=\frac12(\Delta_i+\Delta_j)
\eeq
are also additive. Beyond the hard core, the $ij$ interaction is attractive if $\epsilon_{ij}<0$ and repulsive if $\epsilon_{ij}>0$.

To ensure that no particle of a third species can simultaneously overlap with the interaction shells of particles $i$ and $j$ when their mutual separation lies within the first shell, we impose the geometrical condition $\Delta_{ij}\leq\sigma_{\min}\equiv\min{\sigma_k}$. Under this condition, no three-particle overlap configuration involving a third species can occur when $\sigma_{ij}\leq r\leq\lambda_{ij}$. The RDF $g_{ij}(r)$ in this region can nevertheless be affected indirectly by the other components through many-particle correlations.

We now define the Laplace transform of the product  $rg_{ij}(r)$ as
\beq
G_{ij}(s)=\int_0^\infty dr\, e^{-s r} r g_{ij}(r).
\eeq
The contact value $g_{ij}(\sigma_{ij}^+)$ is given by~\cite{S16}
\beq
\label{2}
g_{ij}(\sigma_{ij}^+)=\sigma_{ij}^{-1}\lim_{s\to\infty}s e^{\sigma_{ij}s}G_{ij}(s),
\eeq
implying the constraint $\lim_{s\to\infty}s e^{\sigma_{ij}s}G_{ij}(s)=\text{finite}$.
Additionally, the condition of finite isothermal compressibility implies the low-$s$ behavior, namely~\cite{YSH98,S16}
\beq
\label{4}
s^2 G_{ij}(s)=1+\mathcal{O}(s^2).
\eeq
The absence of a term linear in $s$ follows from the requirement of finite
isothermal compressibility.

The Fourier transform of the total correlation functions $h_{ij}(r)=g_{ij}(r)-1$ can be obtained from the Laplace transform $G_{ij}(s)$ as follows:
\bal
\widetilde{h}_{ij}(q)=&\frac{4\pi}{q}\int_0^\infty dr\, r\sin(qr)h_{ij}(r)\nn
=&-4\pi\text{Re}\left[\frac{s^2G_{ij}(s)-1}{s^3}\right]_{s=\imath q}.
\eal
Furthermore, the continuity of the cavity function
\beq
y_{ij}(r)\equiv g_{ij}(r)e^{\beta\phi_{ij}(r)}
\eeq
at $r=\lambda_{ij}$, where $\beta\equiv1/k_B T$ ($k_B$ being the Boltzmann constant and $T$ the absolute temperature), implies
\beq
\label{7}
\frac{g_{ij}(\lambda_{ij}^+)}{g_{ij}(\lambda_{ij}^-)}=e^{\beta\epsilon_{ij}}.
\eeq

For completeness, we note that, in the zero-density limit,
\beq
\lim_{\rho\to 0}g_{ij}(r)=\begin{cases}
  0,&r<\sigma_{ij},\\
  e^{-\beta\epsilon_{ij}},&\sigma_{ij}<r<\lambda_{ij},\\
  1,&r>\lambda_{ij}.
\end{cases}
\eeq
Consequently,
\bal
\label{5}
\lim_{\rho\to 0}G_{ij}(s)=&e^{-\beta\epsilon_{ij}}e^{-\sigma_{ij}s}\frac{1+\sigma_{ij}s}{s^2}\nn
&-\left(e^{-\beta\epsilon_{ij}}-1\right)e^{-\lambda_{ij}s}\frac{1+\lambda_{ij}s}{s^2}.
\eal

The narrow-shell limit
$\Delta_{i}\ll\sigma_{i}$
will play a central role in the development below, since in this regime
the interaction shell behaves similarly to an effective sticky layer if $\epsilon_{ij}\leq 0$.
This analogy provides the basis for constructing the present
RFA from the known PY solution of SHS mixtures.

\section{SHS mixtures and  the PY solution}
\label{sec3}
As the present theory is constructed upon the exact PY solution for SHS mixtures, we first summarize the relevant results for such systems.

\subsection{General relations}
\label{sec3A}

The SHS limit corresponds to $\Delta_i\to0$ (so that $\Delta_{ij}\to0$,  $\lambda_{ij}\to\sigma_{ij}$) and
$\epsilon_{ij}\to-\infty$,
while keeping the stickiness parameters
\beq
\label{alphaij}
\alpha_{ij}\equiv (e^{-\beta\epsilon_{ij}}-1)\Delta_{ij}
\eeq
constant.

In this limit, Eqs.~\eqref{2} and \eqref{7} merge into the single condition
\beq
\label{8SHS}
e^{\sigma_{ij}s}G_{ij}^\shs(s)=\sigma_{ij}y_{ij}^\shs(\sigma_{ij})\left(\alpha_{ij}+s^{-1}\right)+\mathcal{O}(s^{-2}).
\eeq
The Boltzmann factor becomes
\beq
e^{-\beta\phi_{ij}(r)}
=
\Theta(r-\sigma_{ij})
+\alpha_{ij}\delta(r-\sigma_{ij}),
\eeq
which implies~\cite{S16}
\beq
\label{8bSHS}
g_{ij}^\shs(r)=y_{ij}^\shs(r)\Theta(r-\sigma_{ij})+\alpha_{ij}y_{ij}^\shs(\sigma_{ij})\delta(r-\sigma_{ij}) .
\eeq
According to Eq.~\eqref{8SHS}, the contact value $y_{ij}^\shs(\sigma_{ij})$ can be obtained from $G_{ij}^\shs(s)$ as
\beq
\label{10SHS}
\sigma_{ij}\alpha_{ij}y_{ij}^\shs(\sigma_{ij})=\lim_{s\to\infty} e^{\sigma_{ij}s}G_{ij}^\shs(s).
\eeq
Equation~\eqref{8SHS} also implies
\beq
\label{9SHS}
\lim_{s\to\infty}s\left[ e^{\sigma_{ij}s}G_{ij}^\shs(s)-\sigma_{ij}\alpha_{ij}y_{ij}^\shs(\sigma_{ij})\right]=\sigma_{ij}y_{ij}^\shs(\sigma_{ij}).
\eeq

If we denote by $Y_{ij}(s)$ the Laplace transform of $ry_{ij}(r)\Theta(r-\sigma_{ij})$, Eq.~\eqref{8bSHS} gives the relationship
\beq
Y_{ij}^\shs(s)=G_{ij}^\shs(s)-\sigma_{ij}\alpha_{ij}y_{ij}^\shs(\sigma_{ij}) e^{-\sigma_{ij}s},
\eeq
so that, in analogy to Eq.~\eqref{2},
\beq
\label{2y}
y_{ij}^\shs(\sigma_{ij})=\sigma_{ij}^{-1}\lim_{s\to\infty}s e^{\sigma_{ij}s}Y_{ij}^\shs(s).
\eeq

\subsection{PY solution}
Within the PY approximation, the Laplace transform
$G_{ij}^{\shs}(s)$ admits the exact representation \cite{PS75,B75,BT79,SYH98}:
\beq
\label{16a}
G_{ij}^\shs(s)=\frac{e^{-\sigma_{ij}s}}{ s^2}\left(\mathsf{N}(s)\cdot\left[\mathsf{I}+\mathsf{B}(s)\right]^{-1}\right)_{ij},
\eeq
where $\mathsf{I}$ is the $n\times n$ identity matrix and
\begin{subequations}
\label{16cb}
\beq
\label{16c}
N_{ij}(s)=\sum_{m=0}^2N_{ij}^{(m)}s^m,
\eeq
\beq
\label{16b}
B_{ij}(s)=\frac{2\pi\rho x_i}{s^3}\sum_{m=0}^2N_{ij}^{(m)}s^m\varphi_{2-m}(\sigma_i s).
\eeq
\end{subequations}
Here, we have introduced the auxiliary functions
\beq
\varphi_m(x)\equiv e^{-x}-\sum_{\ell=0}^m\frac{(-1)^\ell}{\ell!}x^\ell.
\eeq
Note that, according to Eqs.~\eqref{10SHS} and \eqref{9SHS},
\begin{subequations}
\beq
\sigma_{ij}\alpha_{ij}y_{ij}^\shs(\sigma_{ij})=N^{(2)}_{ij},
\eeq
\beq
\label{9bSHS}
\alpha_{ij}\lim_{s\to\infty}s\left[ e^{\sigma_{ij}s}G_{ij}^\shs(s)-N^{(2)}_{ij}\right]=N^{(2)}_{ij}.
\eeq
\end{subequations}

The $3n^2$ coefficients $\{N_{ij}^{(0)}\}$, $\{N_{ij}^{(1)}\}$, and $\{N_{ij}^{(2)}\}$ are determined by imposing the exact low- and high-$s$ constraints given by Eqs.~\eqref{4} and \eqref{9bSHS}, respectively. First, Eq.~\eqref{4} allows one to express $\{N_{ij}^{(0)}\}$ and $\{N_{ij}^{(1)}\}$ in terms of $\{N_{ij}^{(2)}\}$ as \cite{SYH98}
\begin{subequations}
\beq
N_{ij}^{(0)}=c_1 +2c_2\sigma _{j} -c_1\rho \sum_{k}x _{k} \sigma
_{k}N_{kj}^{(2)},
\label{20SHS}
\eeq
\begin{equation}
N_{ij}^{(1)}=c_1 \sigma _{ij}+{c_2}\sigma
_{i}\sigma _{j} -\frac{c_1}{2} \sigma _{i}\rho \sum_{k}x _{k} \sigma
_{k}N_{kj}^{(2)} ,
\label{21SHS}
\end{equation}
\end{subequations}
where
\beq
c_1\equiv\frac{1}{1-\frac{\pi}{6}\rho\langle\sigma^3\rangle},
\quad c_2\equiv\frac{\frac{\pi}{4}\rho\langle\sigma^2\rangle}{\left(1-\frac{\pi}{6}\rho\langle\sigma^3\rangle\right)^2}.
\eeq
Here, $\langle \sigma^\ell\rangle\equiv\sum_k x_k\sigma_k^\ell$.

Finally, Eq.~\eqref{9bSHS} yields the following set of bilinear equations for $\{N_{ij}^{(2)}\}$ \cite{SYH98}:
\bal
N_{ij}^{(2)}=&\alpha_{ij}\left[c_1 \sigma_{ij}+c_2\sigma _{i}\sigma _{j}-\frac{c_1}{2}\rho
\sum_{k}x _{k}\sigma _{k}\left( N_{ki}^{(2)}\sigma
_{j}+N_{kj}^{(2)}\sigma _{i}\right) \right.\nn
&\left.+\rho\sum_{k}x_{k}N_{ki}^{(2)}N_{kj}^{(2)}\right].
\label{26SHS}
\eal
Equation~\eqref{26SHS} constitutes the only nonlinear part of the PY solution and fully determines the contact values of the cavity functions.
The PY solution for AHS mixtures~\cite{L64,YSH98} is recovered by setting
$\alpha_{ij}=0$, which yields
$N_{ij}^{(2)}=0$.

\subsection{Reformulation of the PY-SHS solution by a change of parameters}

For later convenience, and anticipating the extension to finite-width
SW/SS interactions, it is useful to introduce an alternative parameterization
of the coefficients $N_{ij}^{(m)}$. This rewriting is purely formal at this stage,
since it simply amounts to a change of variables and does not introduce any approximation.

Let us rewrite the coefficients $N_{ij}^{(0)}$, $N_{ij}^{(1)}$, and $N_{ij}^{(2)}$ as
\begin{subequations}
\label{25SHS}
\beq
\label{2.15a}
N_{ij}^{(0)}= L_{ij}^{(0)}+\oL_{ij}^{(0)},
\eeq
\beq
N_{ij}^{(1)}= L_{ij}^{(1)}+\oL_{ij}^{(1)}-\oL_{ij}^{(0)}\tilde{\Delta},
\eeq
\beq
N_{ij}^{(2)}= -\oL_{ij}^{(1)}\tilde{\Delta},
\eeq
\end{subequations}
where $\tilde{\Delta}$ is an arbitrary length parameter that will later be identified with  effective shell widths characterizing genuine SW/SS mixtures (see Sec.~\ref{sec4.2}).
Formally, Eqs.~\eqref{25SHS} can be rewritten as
\beq
\label{25SHSbis}
N_{ij}^{(m)}=L_{ij}^{(m)}+\oL_{ij}^{(m)}-\oL_{ij}^{(m-1)}\tilde{\Delta},
\eeq
with the convention $L_{ij}^{(2)}=\oL_{ij}^{(2)}=\oL_{ij}^{(-1)}=0$.

Insertion of Eq.~\eqref{25SHSbis} allows one to rewrite  Eqs.~\eqref{16cb} as
\begin{subequations}
\label{16ccbb}
\beq
\label{16cc}
N_{ij}(s)=\sum_{m=0}^1\left[L_{ij}^{(m)}+(1-s\tilde{\Delta})\oL_{ij}^{(m)}\right]s^m,
\eeq
\bal
\label{16bb}
B_{ij}(s)=&\frac{2\pi\rho x_i}{s^3}
\sum_{m=0}^1s^m\Big\{L_{ij}^{(m)}\varphi_{2-m}(\sigma_i s)+\oL_{ij}^{(m)}[\varphi_{2-m}(\sigma_i s)\nn
&+s\tilde{\Delta} \varphi_{2-m}'(\sigma_i s)]\Big\},
\eal
\end{subequations}
where we have used the identity $\varphi_{m-1}(x)=-\varphi_m'(x)$.

Inserting Eq.~\eqref{16cc} into Eq.~\eqref{16a}, one has
\bal
\label{16SHS}
G_{ij}^\shs(s)=&\frac{e^{-\sigma_{ij}s}}{ s^2}\left(\mathsf{L}(s)\cdot\left[\mathsf{I}+\mathsf{B}(s)\right]^{-1}\right)_{ij}\nn
&+\frac{e^{-\sigma_{ij}s}}{ s^2}(1-s\tilde{\Delta})\left(\mathsf{\oL}(s)\cdot\left[\mathsf{I}+\mathsf{B}(s)\right]^{-1}\right)_{ij},
\eal
where
\beq
\label{16cSHS}
L_{ij}(s)=L_{ij}^{(0)}+L_{ij}^{(1)}s,\quad \oL_{ij}(s)=\oL_{ij}^{(0)}+\oL_{ij}^{(1)}s.
\eeq
Equation~\eqref{16SHS} separates the PY-SHS solution into two contributions,
a decomposition that will prove particularly convenient for constructing the
RFA of finite-width SW/SS mixtures.

\section{An RFA for narrow SW/SS and weakly NAHS mixtures}
\label{sec4}

\subsection{Aim}

In Ref.~\cite{MYS06}, the PY-SHS solution was employed as an effective representation of SW mixtures with narrow wells through the approximation
\beq
\label{22SHS}
y_{ij}(r)\approx y_{ij}^\shs(r).
\eeq

Our aim now is to propose a more advanced approximation in the same spirit as previously done for the monocomponent case~\cite{YS94,S16,HYS16}. Building upon the PY-SHS solution, our nonperturbative approximation is in principle restricted to narrow wells. On the other hand, simply by changing the sign of the potential-energy parameters $\epsilon_{ij}$, it can also be applied to narrow SS mixtures. Moreover, by taking the limit where the shoulder barriers either vanish or diverge, we will be able to account for the case of NAHS mixtures with weak (positive or negative) nonadditivity.

Our strategy consists of two stages.
First, we construct an approximation $g_{ij}^0(r)$ that preserves the essential structure of the PY-SHS solution while incorporating finite interaction widths, in analogy with the monocomponent case~\cite{YS94,S16,HYS16}.
Second, we restore the smoothness properties of the cavity function by introducing a local correction confined to the first shell.

By construction, the first-stage cavity function is continuous at $r=\lambda_{ij}$, that is, $y_{ij}^0(\lambda_{ij}^-)=y_{ij}^0(\lambda_{ij}^+)$. However, there exists an artificial discontinuity of the derivative of the cavity function $y_{ij}^0(r)$ at $r=\lambda_{ij}$, i.e., the difference
\beq
\label{3.2}
\mu_{ij}\equiv \left.\frac{dy_{ij}^{0}(r)}{dr}\right|_{r=\lambda_{ij}^+}-\left.\frac{dy_{ij}^{0}(r)}{dr}\right|_{r=\lambda_{ij}^-},
\eeq
is in general nonzero.

Thus, in the second stage, we refine our proposal by just adding a function $Q_{ij}(r)$, where $Q_{ij}(\lambda_{ij})=0$ and $Q_{ij}'(\lambda_{ij})=\mu_{ij}$, to $y_{ij}^0(r)$ \emph{only} in the first shell $\sigma_{ij}\leq r\leq\lambda_{ij}$.
More specifically,
\begin{subequations}
\label{3.3}
\beq
y_{ij}(r)=y_{ij}^0(r)+\Theta(r-\sigma_{ij})\Theta(\lambda_{ij}-r)Q_{ij}(r),
\eeq
\beq
g_{ij}(r)= g_{ij}^0(r)+\Theta(r-\sigma_{ij})\Theta(\lambda_{ij}-r)e^{-\beta\epsilon_{ij}}Q_{ij}(r).
\eeq
\end{subequations}
Among the infinitely many possible choices satisfying
$Q_{ij}(\lambda_{ij})=0$ and
$Q_{ij}'(\lambda_{ij})=\mu_{ij}$,
we adopt the quadratic form
\begin{equation}
\label{3.4}
Q_{ij}(r)=\mu_{ij}\frac{r}{\lambda_{ij}}(r-\lambda_{ij}),
\end{equation}
which has been found to yield a better numerical performance than the simplest linear choice.

\subsection{Our proposal}
\label{sec4.2}

The key observation is that, for sufficiently narrow shells,
the first-order expansion
\beq
e^{-\sigma_{ij}s}(1-s\Delta_{ij})
=
e^{-\lambda_{ij}s}
+\mathcal O(\Delta_{ij}^2)
\eeq
suggests a direct correspondence between the PY-SHS structure
and a finite-width interaction shell.
Likewise,
\beq
\varphi_m(\sigma_i s)
+s \Delta_i \varphi_m'(\sigma_i s)
=
\varphi_m(\lambda_i s)
+\mathcal O(\Delta_i^2).
\eeq

Thus, rather than deriving a new approximation from scratch,
we use the PY-SHS solution as the analytical backbone upon which finite-width corrections are built.
Specifically, we  introduce the formal change
\beq
\label{32}
e^{-\sigma_{ij} s}(1-s\tilde{\Delta})\to e^{-\lambda_{ij} s}
\eeq
in Eq.~\eqref{16SHS}, where  $\tilde{\Delta}$  has been identified with $\Delta_{ij}$.
Analogously, by identifying $\tilde{\Delta}$ with $\Delta_i$, we make the formal change
\beq
\label{33}
\varphi_{2-m}(\sigma_i s)+s\tilde{\Delta} \varphi_{2-m}'(\sigma_i s)\to \varphi_{2-m}(\lambda_i s)
\eeq
in Eq.~\eqref{16bb}.

The resulting approximation is then
\begin{subequations}
\label{6&8c}
\bal
\label{6}
G_{ij}^0(s)=&\frac{e^{-\sigma_{ij}s}}{ s^2}\left(\mathsf{L}(s)\cdot\left[\mathsf{I}+\mathsf{A}(s)\right]^{-1}\right)_{ij}\nn
&+\frac{e^{-\lambda_{ij}s}}{ s^2}\left(\mathsf{\oL}(s)\cdot\left[\mathsf{I}+\mathsf{A}(s)\right]^{-1}\right)_{ij},
\eal
\beq
\label{8c}
A_{ij}(s)=\frac{2\pi\rho x_i}{s^3}
\sum_{m=0}^1s^m\left[L_{ij}^{(m)}\varphi_{2-m}(\sigma_i s)+\oL_{ij}^{(m)}\varphi_{2-m}(\lambda_i s)\right].
\eeq
\end{subequations}
Equations~\eqref{6&8c} reduce exactly to the monocomponent RFA of Refs.~\cite{YS94,S16,HYS16} if one takes $\sigma_i=\sigma$, $\lambda_i=\lambda$, and $\epsilon_i=\epsilon$ for arbitrary $n$.

Compared with the PY-SHS solution, the present RFA introduces an additional set of $n^2$ coefficients [see Eqs.~\eqref{25SHS}], so that the Laplace transforms \eqref{6&8c} depend on the $4n^2$ parameters
$\{L_{ij}^{(0)},L_{ij}^{(1)},\oL_{ij}^{(0)},\oL_{ij}^{(1)}\}$.
Their determination requires the enforcement of suitable physical consistency conditions, which will be discussed in Sec.~\ref{sub coeff}.
Before addressing the general case, however, it is instructive to note that the coefficients are completely fixed in the low-density limit by the exact condition \eqref{5}. One finds
\begin{subequations}
\label{12}
\beq
\label{3.7a}
\lim_{\rho\to 0} L_{ij}^{(0)}= e^{-\beta\epsilon_{ij}},
\quad
\lim_{\rho\to 0} \overline{L}_{ij}^{(0)}=1-e^{-\beta\epsilon_{ij}},
\eeq
\beq
\lim_{\rho\to 0} L_{ij}^{(1)}= \sigma_{ij}e^{-\beta\epsilon_{ij}},
\quad
\lim_{\rho\to 0} \overline{L}_{ij}^{(1)}= \lambda_{ij}\left(1-e^{-\beta\epsilon_{ij}}\right).
\eeq
\end{subequations}

The resulting formalism preserves the analytical tractability of the PY-SHS solution while extending its applicability to finite-width interactions and weak nonadditivities.

\subsection{RDF within the first two shells}

An important feature of the representation \eqref{6&8c} is that it retains much of the analytical structure of the PY-SHS solution, thereby allowing an explicit determination of the RDF in the first few coordination shells.

In the limit $s\to\infty$, $A_{ij}(s)\sim s^{-1}\to 0$. Thus, application of Eq.~\eqref{2} gives
\beq
g_{ij}^0(\sigma_{ij}^+)=\frac{L_{ij}^{(1)}}{\sigma_{ij}}.
\eeq
Analogously,
\beq
\label{13}
g_{ij}^0(\lambda_{ij}^+)-g_{ij}^0(\lambda_{ij}^-)=\frac{\oL_{ij}^{(1)}}{\lambda_{ij}}.
\eeq

The RDF in the first shell $\sigma_{ij}\leq r\leq\lambda_{ij}$ is obtained from Eqs.~\eqref{6&8c} by retaining only the contribution proportional to $e^{-\sigma_{ij}s}$.
More specifically,
\beq
\label{3.10}
r g_{ij}^0(r)=\xi_{ij}(r-\sigma_{ij}), \quad \sigma_{ij}<r<\lambda_{ij},
\eeq
where $\xi_{ij}(r)$ is the inverse Laplace transform of
\beq
\label{6hat}
\Xi_{ij}(s)=s\left({\mathsf{L}}(s)\cdot\left[s^3\mathsf{I}+\hat{\mathsf{A}}(s)\right]^{-1}\right)_{ij}.
\eeq
Here, $\hat{\mathsf A}(s)$ is obtained by retaining only the algebraic part of
$\mathsf A(s)$, i.e.,
\beq
\label{8hat}
\hat{A}_{ij}(s)=-{2\pi\rho x_i}
\sum_{m=0}^1\sum_{\ell=0}^{2-m}\frac{(-1)^\ell s^{m+\ell}}{\ell!}\left[L_{ij}^{(m)}\sigma_i^\ell+\oL_{ij}^{(m)}\lambda_i^\ell\right].
\eeq

Analogously, in the second shell $\lambda_{ij}\leq r\leq\lambda_{ij}+\sigma_{\min}$  we have
\beq
\label{3.13}
rg_{ij}^0(r)=\xi_{ij}(r-\sigma_{ij})+\oxi_{ij}(r-\lambda_{ij}),\quad \lambda_{ij}<r<\lambda_{ij}+\sigma_{\min},
\eeq
where $\oxi_{ij}(r)$ is the inverse Laplace transform of
\beq
\label{45}
\oXi_{ij}(s)=s\left({\mathsf{\oL}}(s)\cdot\left[s^3\mathsf{I}+\hat{\mathsf{A}}(s)\right]^{-1}\right)_{ij}.
\eeq

The functions $\xi_{ij}(r)$ and $\oxi_{ij}(r)$ can also be expressed in real space by application of the residue theorem.
In general, the determinant of $s^3\mathsf{I}+\hat{\mathsf{A}}(s)$ has the structure of  a  polynomial  of degree $3n$ in $s$. As a consequence, generically there are $3n$ poles  $\{s_\alpha,\alpha=1,\ldots,3n\}$  of the matrix $\left[s^3\mathsf{I}+\hat{\mathsf{A}}(s)\right]^{-1}$. Therefore,
\beq
\label{37}
\xi_{ij}(r)=\sum_{\alpha=1}^{3n}s_\alpha R_{ij}(s_\alpha)e^{s_\alpha r},
\quad \oxi_{ij}(r)=\sum_{\alpha=1}^{3n}s_\alpha \oR_{ij}(s_\alpha)e^{s_\alpha r},
\eeq
where
\begin{subequations}
\label{28ab}
\beq
\label{28a}
\mathsf{R}(s_\alpha)=\mathsf{L}(s_\alpha)\cdot\mathsf{T}(s_\alpha),
\quad \mathsf{\oR}(s_\alpha)=\mathsf{\oL}(s_\alpha)\cdot\mathsf{T}(s_\alpha),
\eeq
\beq
\label{28b}
\mathsf{T}(s_\alpha)=\lim_{s\to s_\alpha}(s-s_\alpha)\left[s^3\mathsf{I}+\hat{\mathsf{A}}(s)\right]^{-1}.
\eeq
\end{subequations}

From Eqs.~\eqref{3.10} and \eqref{3.13} we obtain
\begin{subequations}
\label{42b}
\beq
\frac{d}{dr}\left[ry_{ij}^0(r)\right]_{r=\lambda_{ij}^-}=  \xi_{ij}'(\Delta_{ij})e^{\beta\epsilon_{ij}},
\eeq
\beq
\frac{d}{dr}\left[ry_{ij}^0(r)\right]_{r=\lambda_{ij}^+}=
  \xi_{ij}'(\Delta_{ij})+\oxi_{ij}'(0).
\eeq
\end{subequations}
Thus, the parameter $\mu_{ij}$ defined in Eq.~\eqref{3.2} is
\beq
\label{3.18}
\mu_{ij}=\frac{\oxi_{ij}'(0)-\left(e^{\beta\epsilon_{ij}}-1\right)\xi_{ij}'(\Delta_{ij})}{\lambda_{ij}}.
\eeq
Equation~\eqref{3.18} provides an explicit measure of the spurious lack of smoothness introduced by the approximation.

The derivative $\xi_{ij}'(r)$ can be obtained from Eq.~\eqref{37} or, alternatively, as the inverse Laplace transform of $s\Xi_{ij}(s)-\xi_{ij}(0)$.
As for $\oxi_{ij}'(0)$, one could use Eq.~\eqref{37} again, but it is more straightforward to consider the expansion of $\oXi(s)$ for large $s$,
\beq
\oXi_{ij}(s)=\oxi_{ij}(0)s^{-1}+\oxi_{ij}'(0)s^{-2}+\cdots.
\eeq
yielding
\begin{subequations}
\beq
\oxi_{ij}(0)=\oL_{ij}^{(1)},
\eeq
\bal
\oxi_{ij}'(0)=&\oL_{ij}^{(0)}-2\pi\rho\sum_k\oL_{ik}^{(1)}
 x_k\Bigg(L_{kj}^{(1)}\sigma_k+\oL_{kj}^{(1)}\lambda_k\nn
 &-L_{kj}^{(0)}\frac{\sigma_k^2}{2}-\oL_{kj}^{(0)}\frac{\lambda_k^2}{2}\Bigg).
\eal
\end{subequations}

\subsection{Determination of the coefficients}
\label{sub coeff}
While Eqs.~\eqref{6&8c} provide the explicit $s$-dependence of the Laplace transforms $G_{ij}^0(s)$, it contains $4n^2$ parameters to be determined.

\subsubsection{Low-$s$ consistency condition}

As proved in the Appendix, Eq.~\eqref{4} provides  the following $2n^2$ constraints:
\begin{subequations}
\label{23}
    \bal
    \label{23a}
    L_{ij}^{(0)}+\oL_{ij}^{(0)}=&1+\pi\rho\sum_k x_k\left[L_{kj}^{(1)}\sigma_k^2+\oL_{kj}^{(1)}\lambda_k^2-L_{kj}^{(0)}\frac{\sigma_k^3}{3}\right.\nn
    &\left.-\oL_{kj}^{(0)}\frac{\lambda_k^3}{3}\right],
    \eal
    \bal
    \label{23b}
    L_{ij}^{(1)}+\oL_{ij}^{(1)} =&\sigma_{ij}+\oX_{ij}^{(0)}\Delta_{ij}+\pi\rho\sum_k x_k\left(\sigma_{ik}+\oX_{ik}^{(0)}\Delta_{ik}\right)\nn
    &\times\Bigg[L_{kj}^{(1)}\sigma_k^2+\oL_{kj}^{(1)}\lambda_k^2-L_{kj}^{(0)}\frac{\sigma_k^3}{3}-\oL_{kj}^{(0)}\frac{\lambda_k^3}{3}\Bigg]\nn
         & +\frac{\pi}{3}\rho\sum_k x_k\Bigg[L_{kj}^{(0)}\frac{\sigma_k^4}{4}+\oL_{kj}^{(0)}\frac{\lambda_k^4}{4}
            -    L_{kj}^{(1)}\sigma_k^3\nn
            &-\oL_{kj}^{(1)}\lambda_k^3\Bigg],
    \eal
     \end{subequations}
where the matrix $\oX_{ij}^{(0)}$ is given by Eq.~\eqref{37b}.

Equation~\eqref{23b} is substantially simplified  if the well/shoulder widths $\Delta_i=\Delta$ are common to all the components. In that case, Eq.~\eqref{23b} becomes (see the Appendix)
\bal
    \label{23bbis}
    L_{ij}^{(1)}+\oL_{ij}^{(1)} =&\oL_{ij}^{(0)}\Delta+
    \sigma_{ij}+\pi\rho\sum_k x_k\sigma_{ik}\Bigg[L_{kj}^{(1)}\sigma_k^2+\oL_{kj}^{(1)}\lambda_k^2\nn
    &-L_{kj}^{(0)}\frac{\sigma_k^3}{3}-\oL_{kj}^{(0)}\frac{\lambda_k^3}{3}\Bigg]+\frac{\pi}{3}\rho\sum_k x_k\Bigg[L_{kj}^{(0)}\frac{\sigma_k^4}{4}\nn
         & +\oL_{kj}^{(0)}\frac{\lambda_k^4}{4}
            -    L_{kj}^{(1)}\sigma_k^3-\oL_{kj}^{(1)}\lambda_k^3\Bigg].
    \eal

\subsubsection{Continuity of the cavity functions}
An additional set of $n^2$ constraints is obtained by requiring continuity of $y_{ij}^0(r)$ at $r=\lambda_{ij}$. Combination of Eqs.~\eqref{7}, \eqref{13}, and \eqref{3.10} yields
\beq
\label{27}
\frac{\oL_{ij}^{(1)}}{\xi_{ij}(\Delta_{ij})}=e^{\beta\epsilon_{ij}}-1.
\eeq
While Eqs.~\eqref{23} have an algebraic character, Eq.~\eqref{27} constitutes a set of transcendental equations because, as Eq.~\eqref{37} shows, $\xi_{ij}(\Delta_{ij})$ is a transcendental function of the coefficients through the exponentials $e^{s_\alpha\Delta_{ij}}$.

\subsubsection{Closure assumption}
We still need another independent set of $n^2$ equations. We might enforce continuity of the derivative of $y_{ij}^0(r)$ at $r=\lambda_{ij}$, i.e., according to Eq.~\eqref{3.18}, we could impose the conditions $\oxi_{ij}'(0)=\left(e^{\beta\epsilon_{ij}}-1\right)\xi_{ij}'(\Delta_{ij})$. But this would represent another set of $n^2$ transcendental equations, thus making the proposal too cumbersome, even in the binary case ($n=2$).

Instead, we choose to impose $n^2$ algebraic equations at the cost of sacrificing the continuity of the derivative of  $y_{ij}^0(r)$ at $r=\lambda_{ij}$, which we will reestablish at the end by means of Eqs.~\eqref{3.3} and \eqref{3.4}.

To proceed, we first note that, according to Eq.~\eqref{23a}, the sum $L_{ij}^{(0)}+\oL_{ij}^{(0)}\equiv\Lambda_j$ is independent of the index $i$, in analogy to what happens with the coefficient $N_{ij}^{(0)}$ for SHS mixtures [see Eqs.~\eqref{20SHS} and \eqref{2.15a}]. Additionally, Eq.~\eqref{3.7a} shows that
\beq
\label{28x}
\lim_{\rho\to 0}\frac{\oL_{ij}^{(0)}}{L_{ij}^{(0)}}=e^{\beta\epsilon_{ij}}-1.
\eeq
Guided by its exact validity in the low-density limit
and by the analogous structure of the PY-SHS solution,
we postulate that Eq.~\eqref{28x}
remains approximately valid at finite density.
Therefore, we propose
\beq
\label{28}
L_{ij}^{(0)}=\Lambda_j e^{-\beta\epsilon_{ij}},\quad \oL_{ij}^{(0)}=\Lambda_j\left(1-e^{-\beta\epsilon_{ij}}\right).
\eeq
Insertion of this ansatz into Eq.~\eqref{23a} yields
\beq
\label{30}
\Lambda_j=\frac{1+\pi\rho\sum_k x_k\left[L_{kj}^{(1)}\sigma_k^2+\oL_{kj}^{(1)}\lambda_k^2\right]}{1+\frac{\pi}{3} \rho\sum_k x_k\left[e^{-\beta\epsilon_{kj}}\sigma_k^3+\left(1-e^{-\beta\epsilon_{kj}}\right)\lambda_k^3\right]}.
\eeq

Apart from the first-order substitutions given by Eqs.\ \eqref{32} and \eqref{33}, Eq.~\eqref{28}
constitutes the only additional closure assumption with respect to the PY-SHS solution.

\subsubsection{Summary}

In summary, the $4n^2$ coefficients ${L_{ij}^{(0)},L_{ij}^{(1)},\oL_{ij}^{(0)},\oL_{ij}^{(1)}}$, which completely specify the RDFs within our theory, can be determined as follows~\cite{note_26_07_2}.
First, insertion of Eqs.~\eqref{28} and \eqref{30} into Eq.~\eqref{23b} [or Eq.~\eqref{23bbis} if $\Delta_i=\Delta$] expresses $\mathsf{L}^{(1)}$ in terms of $\mathsf{\oL}^{(1)}$.
The latter matrix is then determined by solving the transcendental equations~\eqref{27}.

Because the coefficients are obtained from asymmetric equations,
the approximation does not automatically guarantee the exact symmetry
$g_{ij}(r)=g_{ji}(r)$.
In practice, however, the asymmetry is found to be very small.
Whenever necessary, we enforce this property through the symmetrization
$g_{ij}(r)\to \frac{1}{2}\left[g_{ij}(r)+g_{ji}(r)\right]$.

\subsection{Application to NAHS mixtures}

By construction, our scheme reduces to the PY solution for AHS mixtures of diameters $\{\sigma_i\}$ if either $\epsilon_{ij}\to 0$  or $\Delta_{i}\to 0$, since in both cases the  stickiness parameters $\alpha_{ij}$ [see Eq.~\eqref{alphaij}] vanish.

A less trivial test corresponds to an SS mixture in the limit $\epsilon_{ij}\to \infty$, in which case the interaction potential in Eq.~\eqref{1} becomes that of an AHS mixture of diameters $\{\lambda_i\}$. In that limit, one can check that our scheme yields $L_{ij}^{(0)}=L_{ij}^{(1)}=0$ and
\begin{subequations}
\beq
\label{20AHS}
\oL_{ij}^{(0)}=\bar{c}_1 +2\bar{c}_2\lambda_{j} ,
\quad
\oL_{ij}^{(1)}=\bar{c}_1 \lambda_{ij}+\bar{c}_2\lambda_{i}\lambda_{j} ,
\eeq
\beq
\bar{c}_1\equiv\frac{1}{1-\frac{\pi}{6}\rho\langle\lambda^3\rangle},
\quad \bar{c}_2\equiv\frac{\frac{\pi}{4}\rho\langle\lambda^2\rangle}{\left(1-\frac{\pi}{6}\rho\langle\lambda^3\rangle\right)^2}.
\eeq
\end{subequations}
This is nothing but  the PY solution for AHS mixtures with diameters $\{\lambda_i\}$.

\begin{table}
\caption{Values of $\Delta_1=\Delta_2=\Delta$, $\sigma_i$, and $\epsilon_{ij}$ of a binary SS mixture equivalent to a binary NAHS mixture of diameters $\hs_{1}$, $\hs_{2}$, and $\hs_{12}\neq\frac{1}{2}(\hs_1+\hs_2)$.\label{tab2}}
\begin{ruledtabular}
\begin{tabular}{ccccc}
Nonadditivity&$\Delta$&$\sigma_i$&$\epsilon_{11}=\epsilon_{22}$&$\epsilon_{12}$\\
\hline
Positive&$\displaystyle{\hs_{12}-\frac{\hs_1+\hs_2}{2}}$&$\hs_i$&$0$&$\infty$\\
Negative&$\displaystyle{\frac{\hs_1+\hs_2}{2}-\hs_{12}}$&$\hs_i-\Delta$&$\infty$&$0$\\
\end{tabular}
\end{ruledtabular}
\end{table}

A particularly appealing feature of the present formalism is that
it naturally encompasses weakly NAHS mixtures as limiting cases.
If we start from a generic SS mixture (i.e., $\epsilon_{ij}\geq0$) and then some of the barriers tend to $0$ while other ones tend to $\infty$,  the SS mixture becomes equivalent to an NAHS mixture.

To fix ideas, suppose a binary NAHS mixture of diameters $\hs_1$, $\hs_2$, $\hs_{12}\neq \frac{1}{2}(\hs_1+\hs_2)$. This system can be made equivalent to a binary SS mixture with the values of $\sigma_1$, $\sigma_2$, $\Delta_1=\Delta_2$, and $\epsilon_{ij}$ shown in Table \ref{tab2} for both positive and negative nonadditivity. It can be checked that, in those limits,
 $L_{12}^{(0)}=\oL_{ii}^{(0)}=\oL_{ii}^{(1)}=0$ for positive nonadditivity, while $L_{ii}^{(0)}=\oL_{12}^{(0)}=\oL_{12}^{(1)}=0$  for negative nonadditivity.

\section{Comparison with Monte Carlo simulations}
\label{sec5}

In this section we assess the performance of the present RFA by comparing its predictions with Monte Carlo (MC) simulation data for a variety of representative systems.
For simplicity, we restrict ourselves in the main text to symmetric binary mixtures
($x_1=x_2=\frac{1}{2}$,
$\sigma_1=\sigma_2=\sigma_{12}=1$)
with either
$\epsilon_{11}=\epsilon_{22}=0$
or
$\epsilon_{12}=0$.
In the former case, only the unlike width
$\Delta_{12}$
is relevant, whereas in the latter only
$\Delta_{11}=\Delta_{22}$
plays a role.
Whenever $\epsilon_{ij}\neq0$, we plot the cavity functions
$y_{ij}(r)$ rather than the RDFs in order to avoid the trivial discontinuities associated with the interaction potential.

Because the theory is intended for narrow interaction shells, we focus here on systems with
$\Delta\leq 0.1$.
Additional examples, including a test of the theory for an asymmetric ternary mixture, are presented in the SM~\cite{SM_26}.

\subsection{Binary SW mixtures}

\begin{figure}[htbp]
      \includegraphics[height=\figureheight]{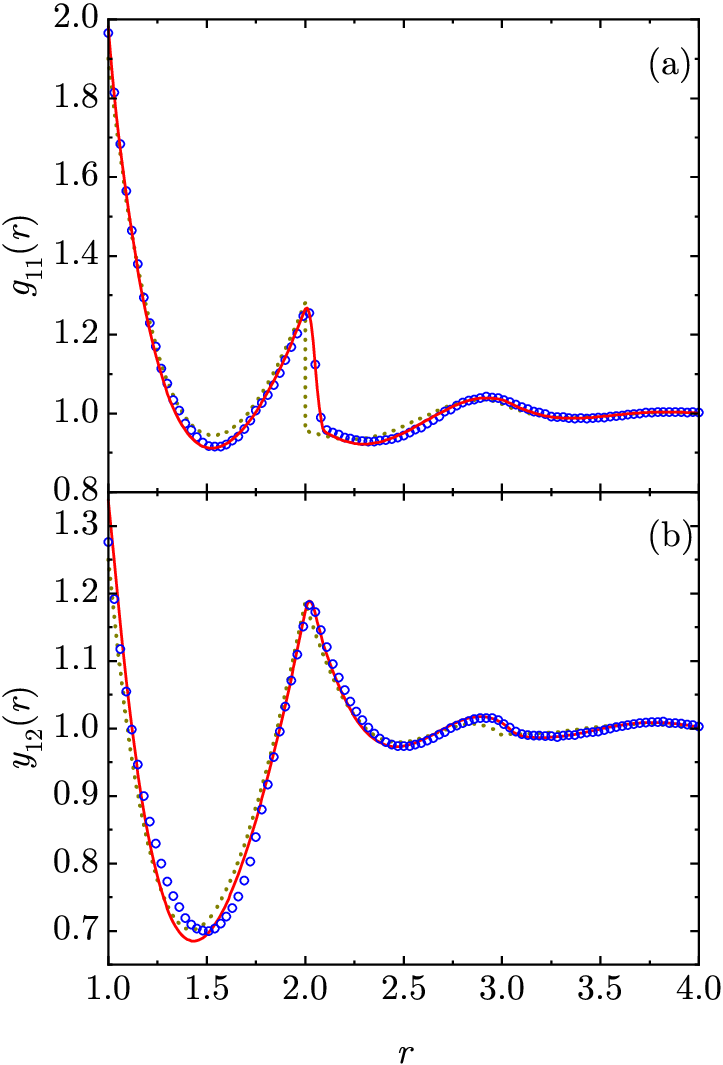}
            \caption{
            Plot of (a) $g_{11}(r)$ and (b) $y_{12}(r)$ for a binary SW mixture with $x_1=\frac{1}{2}$, $\sigma_1=\sigma_2=1$, $\epsilon_{11}=\epsilon_{22}=0$, $\beta\epsilon_{12}=-2.234$, $\Delta_{12}=0.05$, and  $\rho=0.764$. The symbols are MC data \cite{MYS06}, the solid lines represent our RFA results, and the dotted lines represent the PY-SHS prediction with the same value of the stickiness parameters:  $\alpha_{11}=\alpha_{22}=0$, $\alpha_{12}=(e^{-\beta\epsilon_{12}}-1)\Delta_{12}=0.417$.
   \label{fig11_Alex}}
\end{figure}

\begin{figure}[htbp]
      \includegraphics[height=\figureheight]{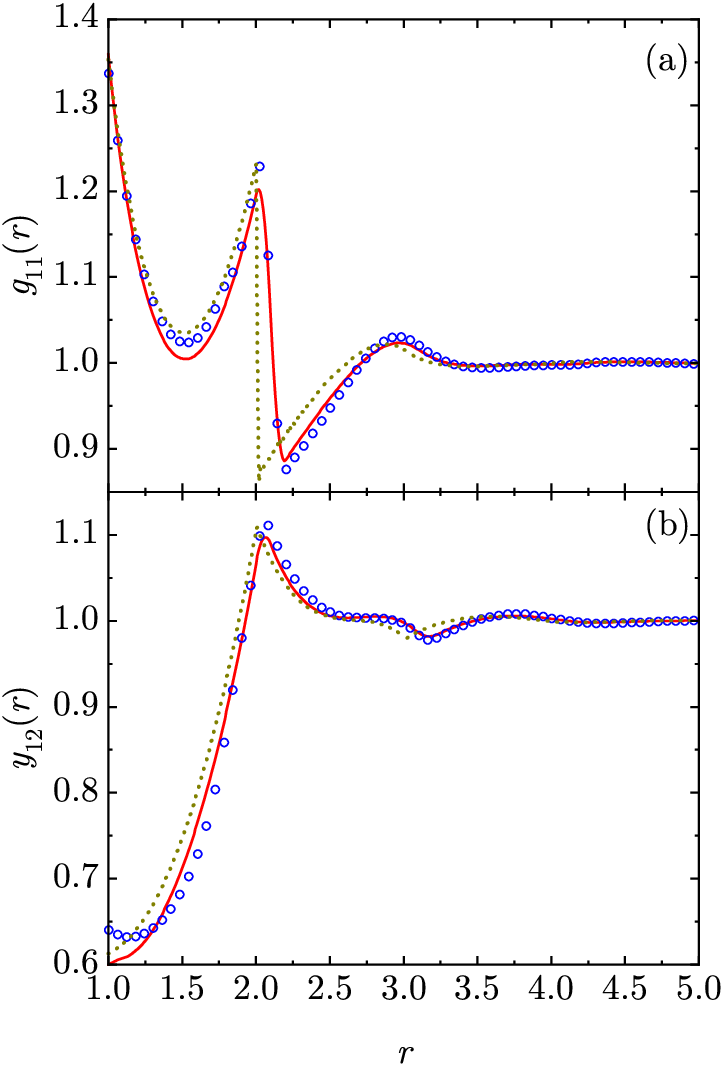}
            \caption{
            Plot of (a) $g_{11}(r)$ and (b) $y_{12}(r)$ for a binary SW mixture with $x_1=\frac{1}{2}$, $\sigma_1=\sigma_2=1$, $\epsilon_{11}=\epsilon_{22}=0$, $\beta\epsilon_{12}=-2.5$, $\Delta_{12}=0.1$, and  $\rho=0.5$. The symbols are MC data \cite{Gianmarco26}, the solid lines represent our RFA results, and the dotted lines represent the PY-SHS prediction with the same value of the stickiness parameters:  $\alpha_{11}=\alpha_{22}=0$, $\alpha_{12}=(e^{-\beta\epsilon_{12}}-1)\Delta_{12}=1.118$.
  \label{Delta0p1_rho0p5}}
\end{figure}

We first present comparisons with MC simulation data \cite{MYS06,Gianmarco26} for symmetric binary mixtures where only the unlike interaction is of SW type. Figure~\ref{fig11_Alex} shows an example with $\Delta_{12}=0.05$, while one with a wider well, $\Delta_{12}=0.1$, is considered in Fig.~\ref{Delta0p1_rho0p5}. A case similar to that of Fig.~\ref{fig11_Alex}, except that $\beta\epsilon_{12}=-1.466$, is shown as Fig.~\SOne
of the SM~\cite{SM_26}. Likewise, Figs.~\STwo, \SThree,  \SFour, and \SFive\,  of the SM display cases similar to that of Fig.~\ref{Delta0p1_rho0p5}, except that $\rho=0.1$, $0.2$, $0.3$, and $0.4$, respectively.

Overall, the RFA provides a satisfactory description of both structural functions.
In particular, it clearly improves upon the PY-SHS approximation by removing the spurious discontinuity exhibited by $g_{11}(r)$ at $r=2$ and by yielding a more accurate representation of the oscillatory structure.
On the other hand, both approaches tend to underestimate the contact values.

\begin{figure}
      \includegraphics[height=\figureheight]{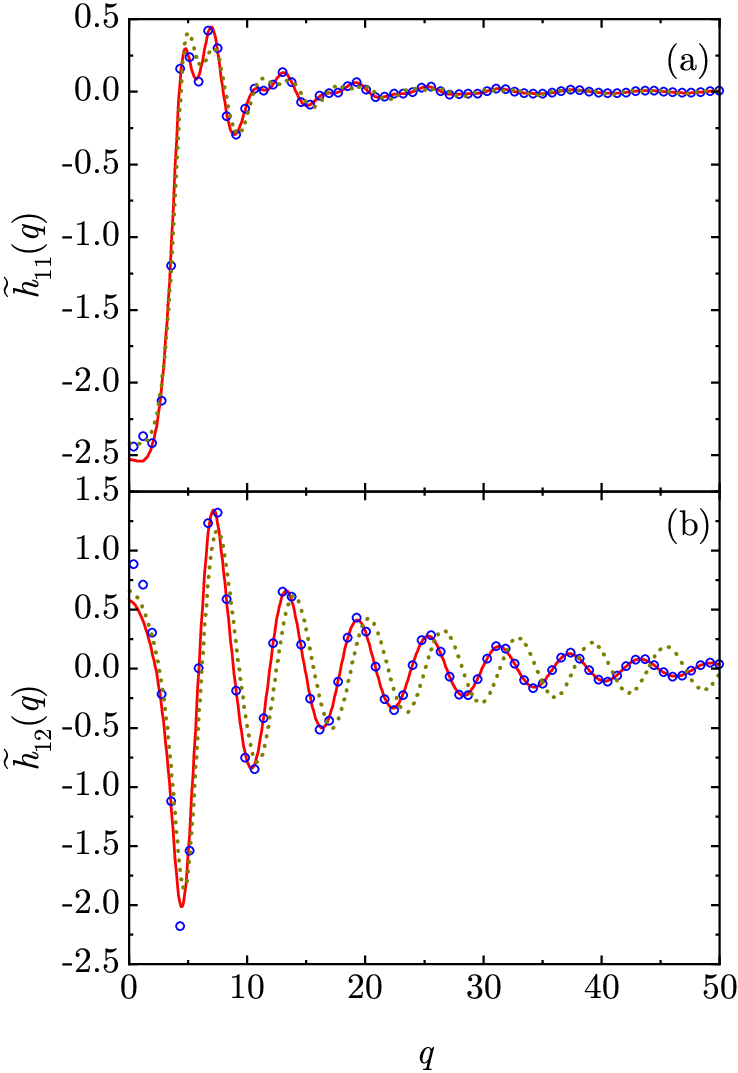}
            \caption{Plot of (a) $\widetilde{h}_{11}(q)$ and (b) $\widetilde{h}_{12}(q)$ for the same system as in Fig.~\ref{Delta0p1_rho0p5}.
  \label{htilde_Delta0p1_rho0p5}}
\end{figure}

The Fourier transforms $\widetilde{h}_{ij}(q)$ for the case of Fig.~\ref{Delta0p1_rho0p5} are shown in Fig.~\ref{htilde_Delta0p1_rho0p5}.
See also Figs.~\SSix--\SNine\, of the SM for the cases of Figs.~\STwo--\SFive.
The agreement in Fourier space is noticeably better than in real space.
This suggests that the residual errors of the approximation are rather local and tend to be smoothed out upon Fourier transformation.

Figure~\ref{htilde_Delta0p1_rho0p5} provides further insight into the different performance of the PY-SHS approximation for the two correlation functions. For $\widetilde{h}_{11}(q)$ [Fig.~\ref{htilde_Delta0p1_rho0p5}(a)], the PY-SHS prediction remains reasonably close to the RFA and MC results over the range of wave numbers considered. In contrast, for $\widetilde{h}_{12}(q)$ [Fig.~\ref{htilde_Delta0p1_rho0p5}(b)], the agreement is good at small $q$ but progressively deteriorates as $q$ increases, with the PY-SHS oscillations eventually becoming out of phase with the RFA and MC results. This difference can be understood from the nature of the corresponding interactions. While the $1$--$1$ interaction is purely HS, the $1$--$2$ interaction contains an attractive shell of finite width $\Delta_{12}$. The PY-SHS approximation replaces this finite-range shell by an effective sticky contact, preserving its integrated effect through the stickiness parameter $\alpha_{12}$. This mapping can provide a reasonable description of the long-wavelength correlations, explaining the good agreement at small $q$, where the Fourier transform is rather insensitive to the detailed spatial extent of the shell.
However, as $q$ increases, the relevant dimensionless parameter $q\Delta_{12}$ becomes progressively larger, and the Fourier transform becomes increasingly sensitive to the finite width of the attractive shell. In the present case, $\Delta_{12}=0.1$, so that $q\Delta_{12}$ reaches values of about $5$ over the range displayed in Fig.~\ref{htilde_Delta0p1_rho0p5}. The replacement of the finite-width shell by a sticky contact can then produce an increasingly significant phase difference in the oscillatory Fourier transform, explaining the out-of-phase behavior of the PY-SHS prediction in Fig.~\ref{htilde_Delta0p1_rho0p5}(b).

Although the RFA theory has been constructed for narrow shells, it is instructive to test its robustness at the extreme value
$\Delta_{12}=\sigma_{\min}$.
This is done in Figs.~\STen\,  and \SEleven\, of the SM~\cite{SM_26}.
As expected, the agreement deteriorates and becomes essentially qualitative.
Nevertheless, the RFA still captures the overall shape and oscillatory behavior of the correlation functions, whereas the PY-SHS approximation becomes much less reliable.
Again, the behavior of the RFA appears as more reliable in the Fourier representation (see Figs.~\STwelve\,  and \SThirteen\,  of the SM).

\subsection{Binary SS mixtures}

We next consider repulsive SS interactions. As anticipated from the construction of the theory, changing the sign of the interaction energies leaves its formal structure unchanged, although its quantitative performance may differ from that for SW interactions.

\begin{figure}
      \includegraphics[height=\figureheight]{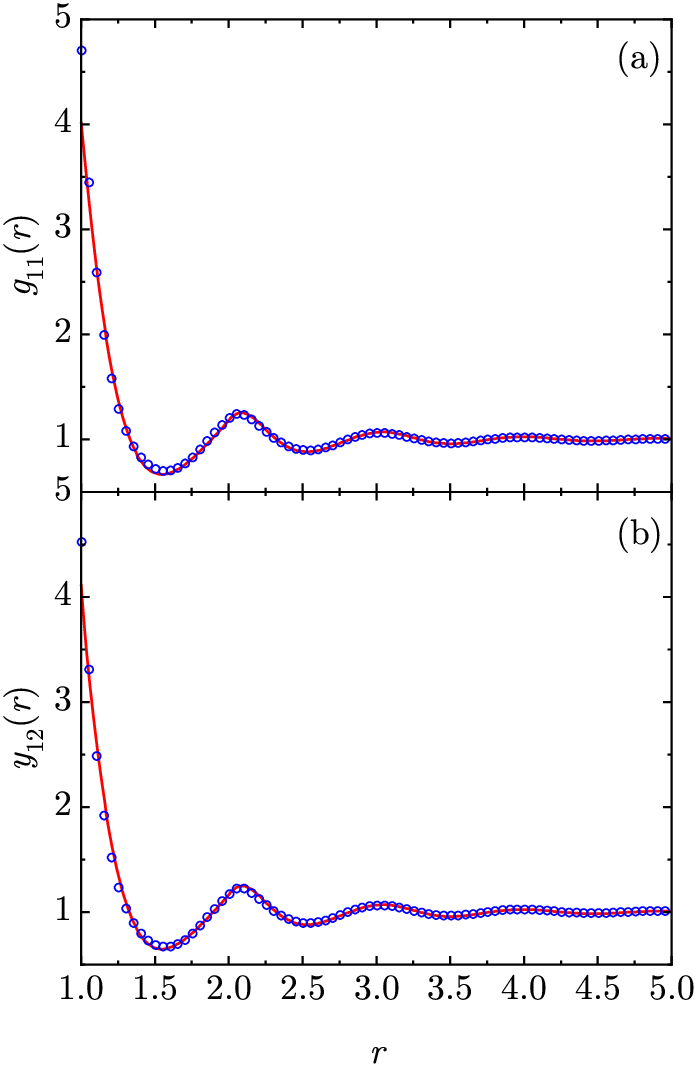}
            \caption{
            Plot of (a) $g_{11}(r)$ and (b) $y_{12}(r)$ for a binary SS mixture with $x_1=\frac{1}{2}$, $\sigma_1=\sigma_2=1$, $\epsilon_{11}=\epsilon_{22}=0$, $\beta\epsilon_{12}=1$, $\Delta_{12}=0.05$, and  $\rho=0.8$. The symbols are our own MC data and the solid lines represent our RFA results.
  \label{SS12_Delta0p05_rho0p8}}
\end{figure}

\begin{figure}
      \includegraphics[height=\figureheight]{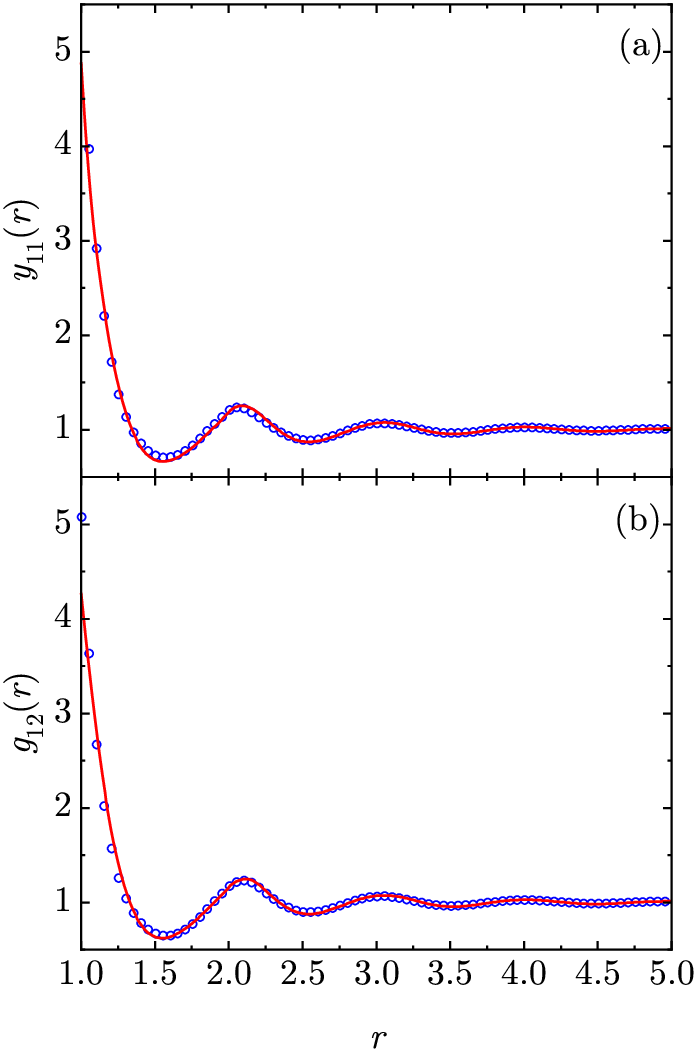}
            \caption{
            Plot of (a) $y_{11}(r)$ and (b) $g_{12}(r)$ for a binary SS mixture with $x_1=\frac{1}{2}$, $\sigma_1=\sigma_2=1$, $\beta\epsilon_{11}=\beta\epsilon_{22}=1$, $\epsilon_{12}=0$, $\Delta_{11}=0.1$, and  $\rho=0.8$. The symbols are our own MC data and the solid lines represent our RFA results.
  \label{SS11_Delta0p1_rho0p8}}
\end{figure}

Figure~\ref{SS12_Delta0p05_rho0p8} shows $g_{11}(r)$ and $y_{12}(r)$ for a symmetric binary mixture in which only the unlike interaction has a repulsive shoulder, with width $\Delta_{12}=0.05$. At $\rho=0.8$ and $\beta\epsilon_{12}=1$, the RFA results agree very well with the MC data. This is complemented by Fig.~\ref{SS11_Delta0p1_rho0p8}, where the like interactions instead exhibit a repulsive shoulder of width $\Delta_{11}=0.1$. Except near contact, where the theoretical results tend to underestimate the MC data, the agreement is again very good.

Interestingly, the four curves shown in Figs.~\ref{SS12_Delta0p05_rho0p8} and \ref{SS11_Delta0p1_rho0p8} practically overlap. In the former case, the like-particle interaction is purely hard-sphere, so that $g_{11}(r)=y_{11}(r)$, whereas in the latter case it is the cross interaction that is purely hard-sphere, so that $g_{12}(r)=y_{12}(r)$. The small differences between the corresponding RDFs arise from the discontinuities associated with the repulsive shoulders: $g_{12}(r)$ in the case of Fig.~\ref{SS12_Delta0p05_rho0p8} and $g_{11}(r)$ in the case of Fig.~\ref{SS11_Delta0p1_rho0p8}. Thus, the cavity functions of these two systems are remarkably similar despite the different widths and locations of the shoulders, much in the spirit of Eq.~\eqref{22SHS}. However, this similarity becomes less pronounced as the shoulder widths increase, as illustrated by Figs.~\SFourteen\, and \SFifteen\, of the SM.

\subsection{Binary NAHS mixtures}

\begin{figure}
      \includegraphics[height=\figureheight]{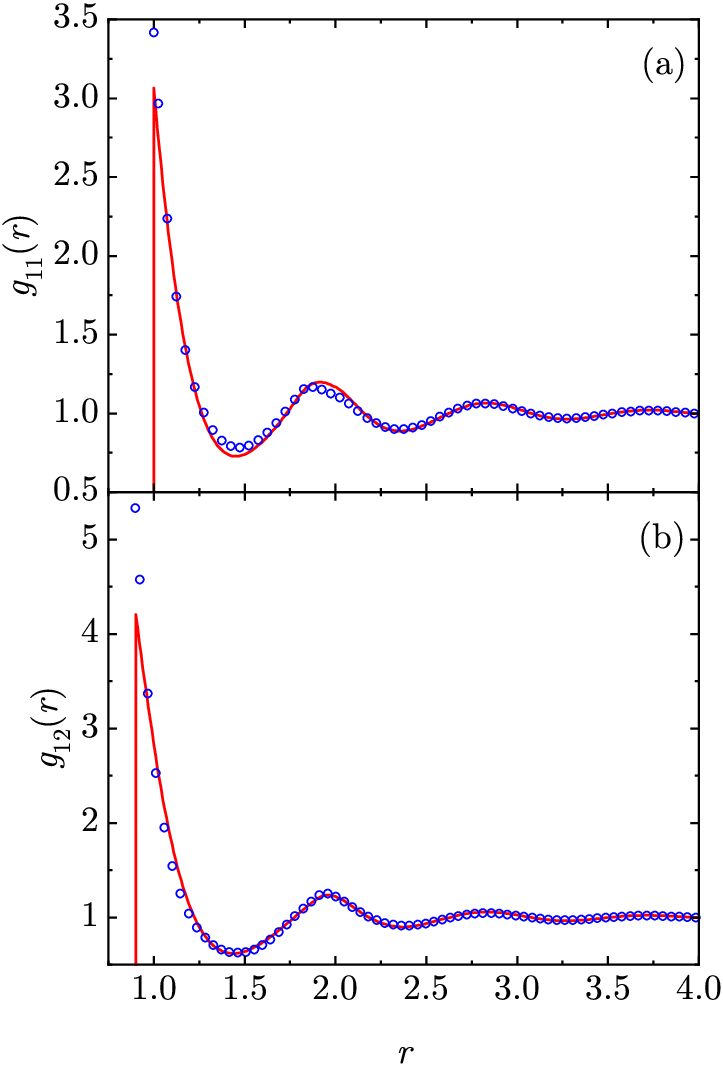}
            \caption{
            Plot of (a) $g_{11}(r)$ and (b) $g_{12}(r)$ for a binary NAHS mixture with $x_1=\frac{1}{2}$, $\hs_1=\hs_2=1$, $\hs_{12}=0.9$, and  $\rho=1$. The symbols are MC data \cite{FS11} and the solid lines represent our RFA results obtained with $\beta\epsilon_{11}=\beta\epsilon_{22}=10$ and $\beta\epsilon_{12}=0$.
            The value $\beta\epsilon_{11}=\beta\epsilon_{22}=10$ is sufficiently large for the results to be effectively indistinguishable from the NAHS limit; larger values produce no appreciable changes in the results shown.
  \label{fig5_Riccardo}}
\end{figure}

\begin{figure}
      \includegraphics[height=\figureheight]{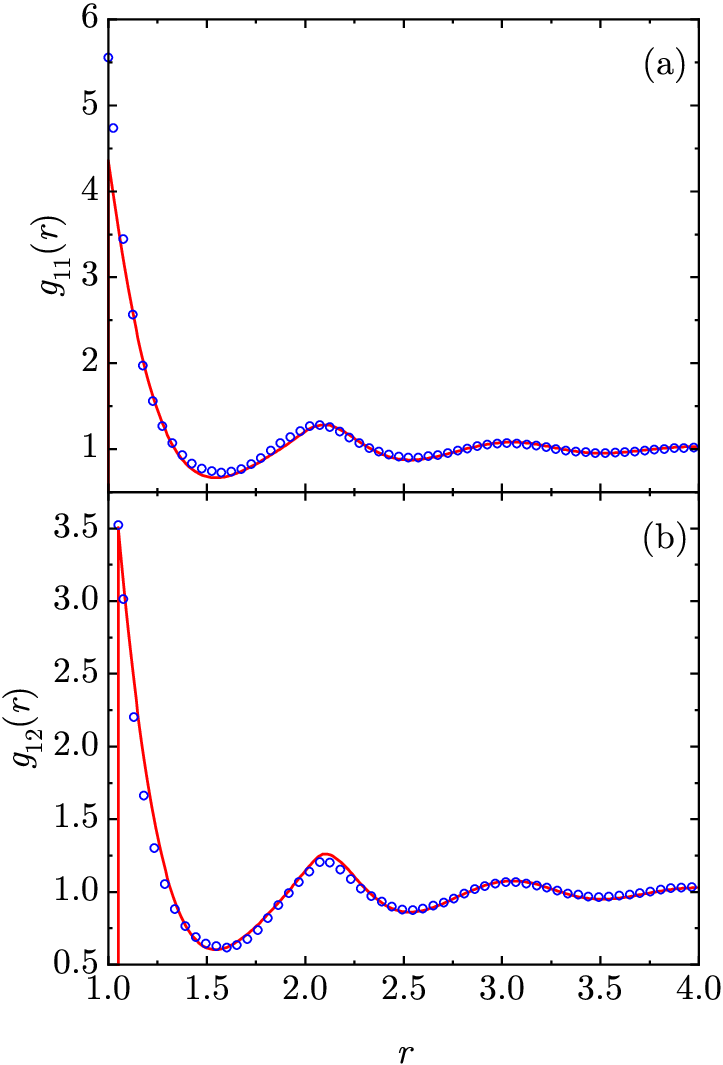}
            \caption{
            Plot of (a) $g_{11}(r)$ and (b)  $g_{12}(r)$ for a binary NAHS mixture with $x_1=\frac{1}{2}$, $\hs_1=\hs_2=1$, $\hs_{12}=1.05$,   and $\rho=0.8$. The symbols are MC data \cite{FS11} and the solid lines represent our RFA results obtained with $\beta\epsilon_{11}=\beta\epsilon_{22}=0$ and $\beta\epsilon_{12}=10$.
            The value $\beta\epsilon_{12}=10$ is sufficiently large for the results to be effectively indistinguishable from the NAHS limit; larger values produce no appreciable changes in the results shown.
  \label{fig8_Riccardo}}
\end{figure}

We next assess the ability of the theory to describe weakly NAHS mixtures, which constitute a nontrivial limiting case of the present formalism.
Representative examples with weak negative and positive nonadditivity are shown in Figs.~\ref{fig5_Riccardo} and \ref{fig8_Riccardo}.
Figure~\ref{fig5_Riccardo} corresponds to a weakly negative nonadditivity, while Fig.~\ref{fig8_Riccardo} corresponds to a weakly positive nonadditivity.

A generally satisfactory agreement is observed in both cases.
The largest discrepancies occur in the vicinity of contact, a feature inherited from the PY description of AHS mixtures \cite{MMYSH02}.

As the magnitude of the nonadditivity increases, the quality of the approximation progressively deteriorates, particularly for the unlike correlation function $g_{12}(r)$ (see Figs.~\SSixteen--\SNineteen\, of the SM).
This behavior is consistent with the fact that the mapping onto an equivalent SS mixture becomes less accurate as the effective shell width increases.

Nevertheless, even outside its nominal regime of applicability, the present approach performs comparably to the alternative RFA specifically devised for NAHS mixtures in Ref.~\cite{FS11}.
This is remarkable in view of the fact that the present theory was not designed primarily for nonadditive systems but rather emerges naturally from the finite-width extension of the PY-SHS solution.

Overall, the comparisons presented in this section indicate that the proposed RFA provides a reliable description of narrow SW and SS mixtures and remains qualitatively accurate even beyond its nominal range of applicability.
Moreover, its ability to reproduce the structure of weakly NAHS mixtures demonstrates the versatility of the formalism and highlights the usefulness of the PY-SHS solution as an analytical starting point for more general interactions.

\section{Concluding remarks}
\label{sec6}

In this paper, we have developed a semianalytical RFA for the structural properties of multicomponent fluids with SW, SS, and weakly NAHS interactions. Our starting point has been the exact analytical structure of the PY solution for additive SHS mixtures. By introducing a formal parameter representing the interaction widths and suitably reinterpreting the SHS solution, we have constructed an extension to finite-width interactions that preserves much of the analytical simplicity of the original PY-SHS theory.

The resulting approximation exhibits a symmetric structure with respect to the hard-core diameters $\{\sigma_i\}$ and the interaction ranges $\{\lambda_i\}$. Its $4n^2$ coefficients are determined by enforcing the exact low-density limit together with a set of physical consistency conditions, including the continuity of the cavity functions at the edges of the interaction shells. The bare RFA yields cavity functions whose derivatives are not exactly continuous at $r=\lambda_{ij}$; this shortcoming is corrected through the introduction of a simple local quadratic contribution confined to the first shell.

Despite its nonperturbative character, the present formulation is expected to be reliable only for sufficiently narrow wells or shoulders and for weak nonadditivities. Comparison with available MC simulations, together with additional simulations performed in this work, shows that the theory provides a generally good description of both real-space and Fourier-space structural properties within its expected domain of applicability. In particular, the RFA systematically improves upon the simpler PY-SHS mapping for narrow SW mixtures. As expected, the quality of the predictions deteriorates as the interaction widths or nonadditivities increase, although the theory often remains qualitatively correct beyond its nominal range of validity.

A limitation inherited from the PY-SHS framework is the tendency to underestimate the contact values of the RDFs. Nevertheless, the overall agreement with simulation data is satisfactory and comparable to that obtained from alternative approaches specifically designed for NAHS mixtures.

Most of the examples considered here correspond to equimolar binary mixtures, while additional tests for asymmetric compositions and ternary mixtures are reported in the SM~\cite{SM_26}. A more extensive assessment of the theory for arbitrary compositions, size asymmetries, and interaction parameters, as well as possible extensions to wider interaction ranges, remains desirable.

More generally, the present work illustrates how the analytically tractable PY solution for SHS mixtures can serve as a useful starting point for constructing semianalytical theories of fluids with finite-range interactions.

\section*{Acknowledgments}
The authors acknowledge financial support from Grant No.~PID2024-156352NB-I00 funded by MCIU/AEI/10.13039/501100011033 and by ERDF/EU, and from Grant No.~GR24022 funded by the Junta de Extremadura (Spain).
The authors are indebted to Gianmarco Muna\`o for carrying out and sharing the MC simulations reported in Figs.~\ref{Delta0p1_rho0p5} and \ref{htilde_Delta0p1_rho0p5} of the main text and Figs.~\STwo--\SThirteen\, of the SM.

\section*{Data availability}
Data supporting figures and numerical results are available from Ref.~\cite{note_26_07_2}.

\appendix

\renewcommand{\theequation}{A\arabic{equation}}
\renewcommand{\theHequation}{A\arabic{equation}}  
\setcounter{equation}{0}

\section*{Appendix: Constraints from Eq.~\eqref{4}}
\label{app:2n2}
Let us start by rewriting Eq.~\eqref{6} as
\beq
\label{B1}
s^2 G_{ij}^0(s)=e^{-\sigma_{ij}s}X_{ij}(s)+e^{-\lambda_{ij}s}\oX_{ij}(s),
\eeq
where
\beq
\label{35}
\mathsf{X}(s)=\mathsf{L}(s)\cdot\left[\mathsf{I}+\mathsf{A}(s)\right]^{-1},\quad \mathsf{\oX}(s)=\mathsf{\oL}(s)\cdot\left[\mathsf{I}+\mathsf{A}(s)\right]^{-1}.
\eeq
The small-$s$ behaviors of $\mathsf{X}(s)$, $\mathsf{\oX}(s)$, and $\mathsf{A}(s)$ are
\begin{subequations}
\beq
\mathsf{X}(s)=\mathsf{X}^{(0)}+\mathsf{X}^{(1)}s+\mathcal{O}(s^2),
\eeq
\beq
\mathsf{\oX}(s)=\mathsf{\oX}^{(0)}+\mathsf{\oX}^{(1)}s+\mathcal{O}(s^2),
\eeq
\beq
\mathsf{A}(s)=\mathsf{A}^{(0)}+\mathsf{A}^{(1)}s+\mathcal{O}(s^2),
\eeq
\end{subequations}
with
\begin{subequations}
\beq
\label{37a}
\mathsf{X}^{(0)}=\mathsf{L}^{(0)}\cdot \left[\mathsf{I}+\mathsf{A}^{(0)}\right]^{-1},
\eeq
\beq
\label{37b}
\mathsf{\oX}^{(0)}=\mathsf{\oL}^{(0)}\cdot \left[\mathsf{I}+\mathsf{A}^{(0)}\right]^{-1},
\eeq
\beq
\label{37c}
\mathsf{X}^{(1)}=\left[\mathsf{L}^{(1)}-\mathsf{X}^{(0)}\cdot\mathsf{A}^{(1)}\right]\cdot \left[\mathsf{I}+\mathsf{A}^{(0)}\right]^{-1},
\eeq
\beq
\label{37d}
\mathsf{\oX}^{(1)}=\left[\mathsf{\oL}^{(1)}-\mathsf{\oX}^{(0)}\cdot\mathsf{A}^{(1)}\right]\cdot \left[\mathsf{I}+\mathsf{A}^{(0)}\right]^{-1},
\eeq
\beq
A_{ij}^{(0)}=\pi\rho x_i\left[L_{ij}^{(1)}\sigma_i^2+\oL_{ij}^{(1)}\lambda_i^2-L_{ij}^{(0)}\frac{\sigma_i^3}{3}-\oL_{ij}^{(0)}\frac{\lambda_i^3}{3}\right],
\eeq
\beq
A_{ij}^{(1)}=\frac{\pi}{3}\rho x_i\left[L_{ij}^{(0)}\frac{\sigma_i^4}{4}+\oL_{ij}^{(0)}\frac{\lambda_i^4}{4}-L_{ij}^{(1)}{\sigma_i^3}-\oL_{ij}^{(1)}{\lambda_i^3}\right].
\eeq
\end{subequations}

Expanding
$e^{-\sigma_{ij}s}$
and
$e^{-\lambda_{ij}s}$
in Eq.\ \eqref{B1} to first order in $s$, and comparing with
Eq.\ \eqref{4},
one obtains
\beq
\mathsf{X}^{(0)}+\mathsf{\oX}^{(0)}=\ones,
\quad \mathsf{X}^{(1)}+\mathsf{\oX}^{(1)}=
\bm{\sigma}+\mathbf{\Delta}\circ \mathsf{\oX}^{(0)}.
\eeq
Here, $\ones$ denotes the matrix whose entries are all equal to unity  and $\mathsf{A}\circ\mathsf{B}$ denotes the Hadamard product, that is, $(\mathsf{A}\circ\mathsf{B})_{ij}=A_{ij}B_{ij}$.
Using Eqs.~\eqref{37a}--\eqref{37c}, the constraints become
\begin{subequations}
\label{39ab}
\beq
\label{39a}
\mathsf{L}^{(0)}+\mathsf{\oL}^{(0)}=\ones+\ones\cdot\mathsf{A}^{(0)},
\eeq
\beq
\label{39b}
\mathsf{L}^{(1)}+\mathsf{\oL}^{(1)}=\left[\bm{\sigma}+\mathbf{\Delta}\circ\mathsf{\oX}^{(0)}\right]\cdot\left[\mathsf{I}+\mathsf{A}^{(0)}\right]+\ones\cdot\mathsf{A}^{(1)}.
\eeq
\end{subequations}
A more explicit form is provided by Eqs.~\eqref{23}.
Equation~\eqref{39a} also shows that
$L_{ij}^{(0)}+\oL_{ij}^{(0)}$
is independent of the row index $i$,
a property already mentioned in Sec.~\ref{sub coeff}.

A particularly useful simplification occurs when all species have the same shell width,
$\Delta_i=\Delta$.
In that case, $\mathbf{\Delta}\circ\mathsf{\oX}^{(0)}= \Delta \mathsf{\oX}^{(0)}$ and Eq.~\eqref{39b} reduces to
\beq
\label{39bbis}
\mathsf{L}^{(1)}+\mathsf{\oL}^{(1)}=\mathsf{\oL}^{(0)}\Delta+\bm{\sigma}\cdot\left[\mathsf{I}+\mathsf{A}^{(0)}\right]+
\ones\cdot\mathsf{A}^{(1)}.
\eeq
This is equivalent to Eq.~\eqref{23bbis} of the main text.

\bibliography{C:/AA_D/Dropbox/Mis_Dropcumentos/bib_files/liquid}

\clearpage
\onecolumngrid   
\begingroup
\renewcommand{\thesection}{S\arabic{section}}
\renewcommand{\thetable}{S\arabic{table}}
\renewcommand{\thefigure}{S\arabic{figure}}
\renewcommand{\theequation}{S\arabic{equation}}
\renewcommand{\theHsection}{S\arabic{section}}
\renewcommand{\theHtable}{S\arabic{table}}
\renewcommand{\theHfigure}{S\arabic{figure}}
\renewcommand{\theHequation}{S\arabic{equation}}

\setcounter{section}{0}
\setcounter{equation}{0}
\setcounter{figure}{0}
\setcounter{table}{0}

\begin{center}
{\large\bfseries Supplemental Material to\\
``A Unified Rational-Function Approximation for Square-Well, Square-Shoulder, and Nonadditive Hard-Sphere Mixtures''}\\[1ex]
Andr\'es Santos,$^{1,2}$ Santos B. Yuste,$^{1,2}$ Ana M. Montero,$^{1}$ and Mariano L\'opez de Haro$^{3}$\\[0.5ex]
{\small
$^1$Departamento de F\'isica, Universidad de Extremadura, E-06006 Badajoz, Spain\\
$^2$Instituto de Computaci\'on Cient\'ifica Avanzada (ICCAEx), Universidad de Extremadura, E-06006 Badajoz, Spain\\
$^3$Instituto de Energ\'{\i}as Renovables, U.N.A.M., Privada Xochicalco s/n, Col. Centro, Temixco, Mor.\ 62580, Mexico}
\end{center}

\begin{quote}
This Supplemental Material contains additional figures complementing those presented in the main text. Figures~\ref{fig12_Alex}--\ref{fig1bT1p5_Munao} correspond to binary square-well (SW) mixtures, Figs.~\ref{SS12_Delta0p25_rho0p3} and \ref{SS11_Delta0p5_rho0p3} correspond to binary square-shoulder (SS) mixtures, Figs.~\ref{fig6_Riccardo}--\ref{fig10_Riccardo} correspond to binary nonadditive hard-sphere (NAHS) mixtures, and Fig.~\ref{Ternary} corresponds to a ternary SW mixture.
\end{quote}

\begin{center}

\begin{minipage}[t]{\mpwidth}
\includegraphics[height=\figheight]{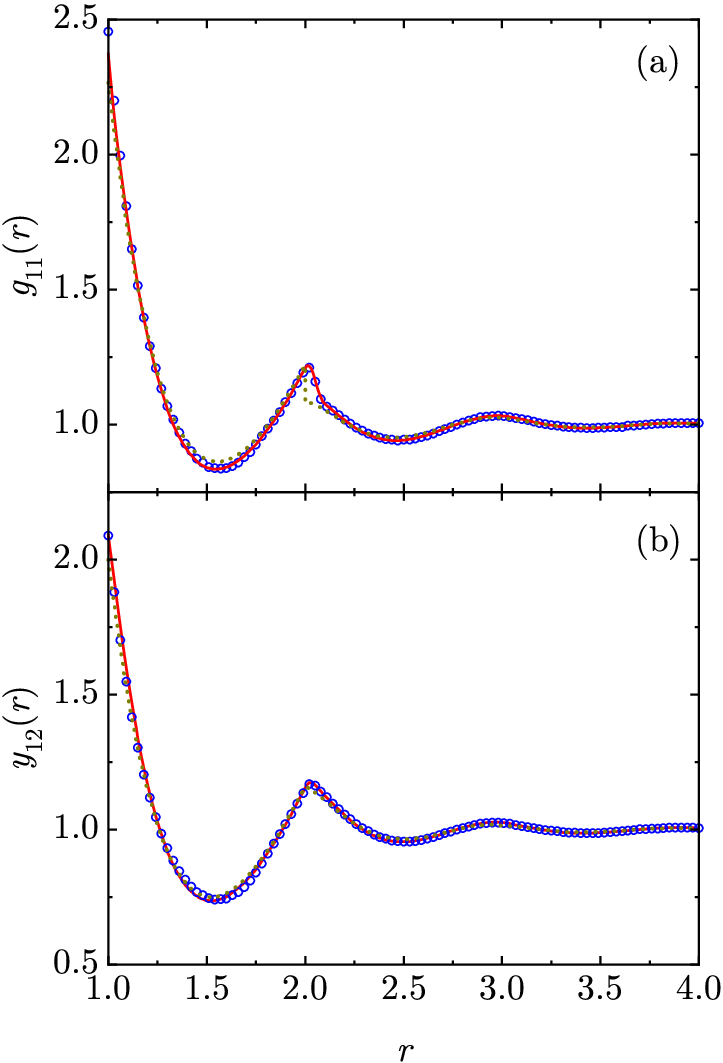}
\captionof{figure}{Plot of (a) $g_{11}(r)$ and (b) $y_{12}(r)$ for a binary SW mixture with $x_1=\frac{1}{2}$, $\sigma_1=\sigma_2=1$, $\epsilon_{11}=\epsilon_{22}=0$, $\beta\epsilon_{12}=-1.466$, $\Delta_{12}=0.05$, and  $\rho=0.764$. The symbols are MC data \cite{MYS06}, the solid lines represent our RFA results, and the dotted lines represent the PY-SHS prediction with the same value of the stickiness parameters:  $\alpha_{11}=\alpha_{22}=0$, $\alpha_{12}=(e^{-\beta\epsilon_{12}}-1)\Delta_{12}=0.167$.}
  \label{fig12_Alex}
\end{minipage}

\begin{minipage}[t]{\mpwidth}
\includegraphics[height=\figheight]{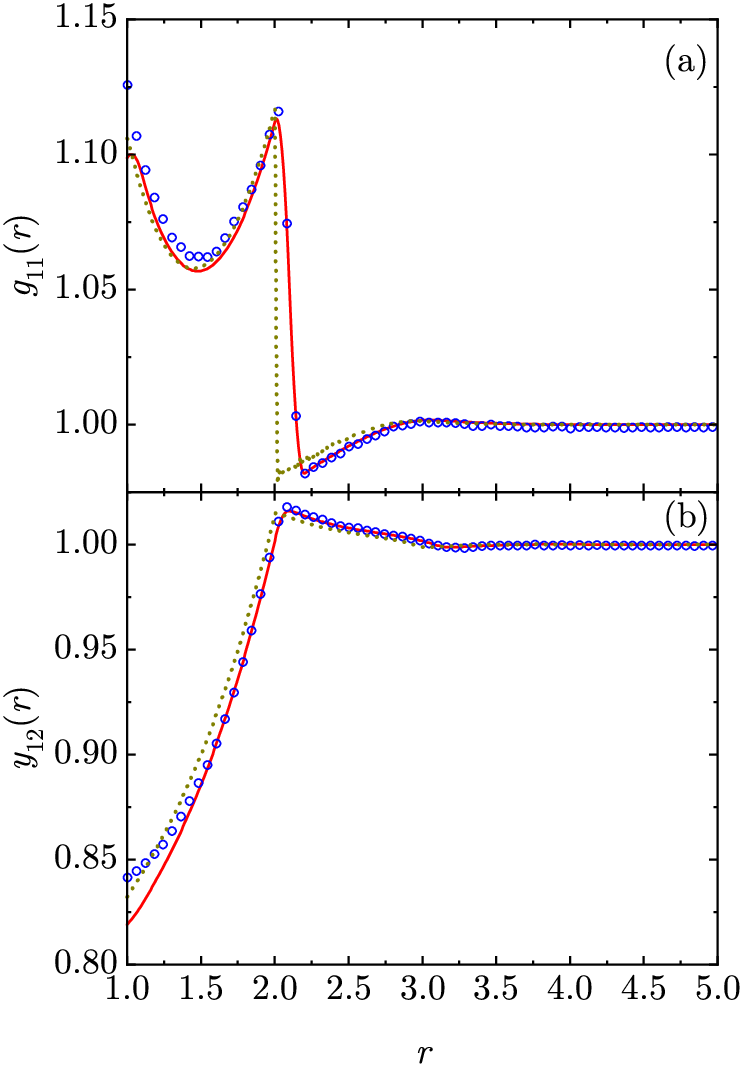}
\captionof{figure}{Plot of (a) $g_{11}(r)$ and (b) $y_{12}(r)$ for a binary SW mixture with $x_1=\frac{1}{2}$, $\sigma_1=\sigma_2=1$, $\epsilon_{11}=\epsilon_{22}=0$, $\beta\epsilon_{12}=-2.5$, $\Delta_{12}=0.1$, and  $\rho=0.1$. The symbols are MC data \cite{Gianmarco26}, the solid lines represent our RFA results, and the dotted lines represent the PY-SHS prediction with the same value of the stickiness parameters:  $\alpha_{11}=\alpha_{22}=0$, $\alpha_{12}=(e^{-\beta\epsilon_{12}}-1)\Delta_{12}=1.118$.
}
\label{Delta0p1_rho0p1}
\end{minipage}
\hfill
\begin{minipage}[t]{\mpwidth}
\includegraphics[height=\figheight]{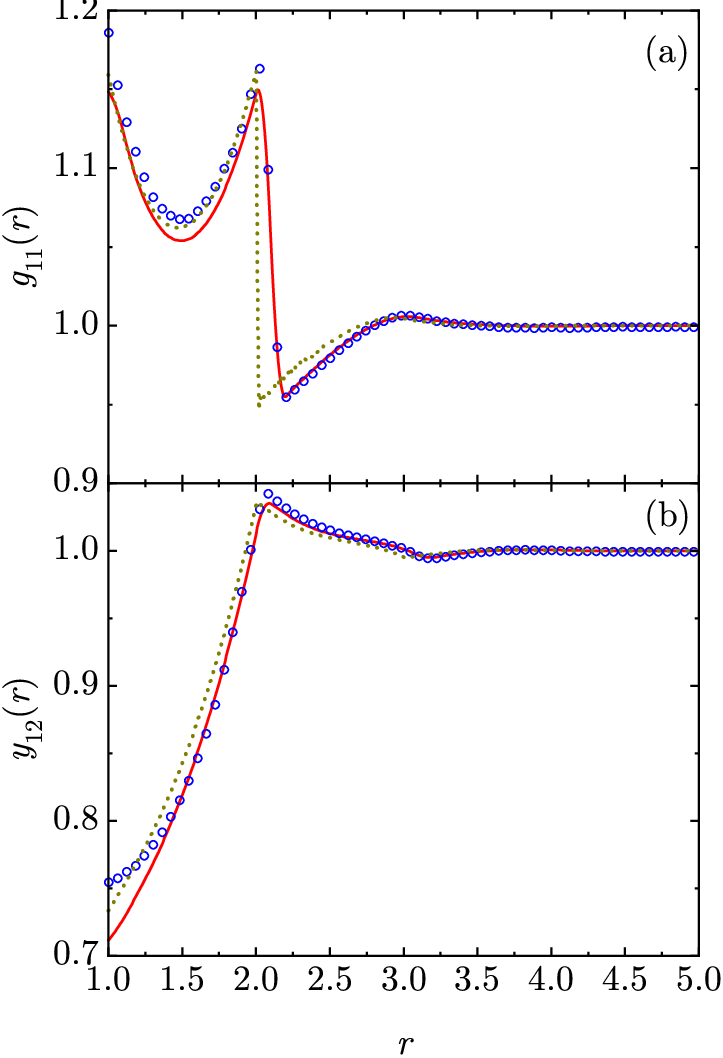}
\captionof{figure}{
Same as in Fig.~\ref{Delta0p1_rho0p1}, except that $\rho=0.2$.
}
\label{Delta0p1_rho0p2}
\end{minipage}

\begin{minipage}[t]{\mpwidth}
\includegraphics[height=\figheight]{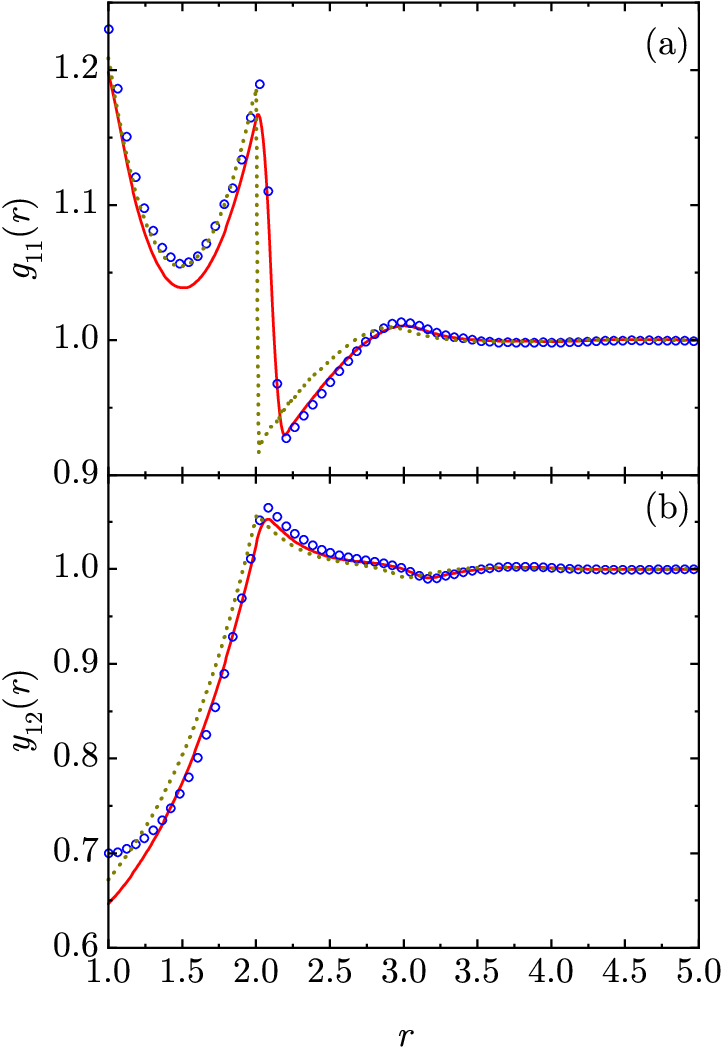}
\captionof{figure}{Same as in Fig.~\ref{Delta0p1_rho0p1}, except that $\rho=0.3$.
}
\label{Delta0p1_rho0p3}
\end{minipage}
\hfill
\begin{minipage}[t]{\mpwidth}
\includegraphics[height=\figheight]{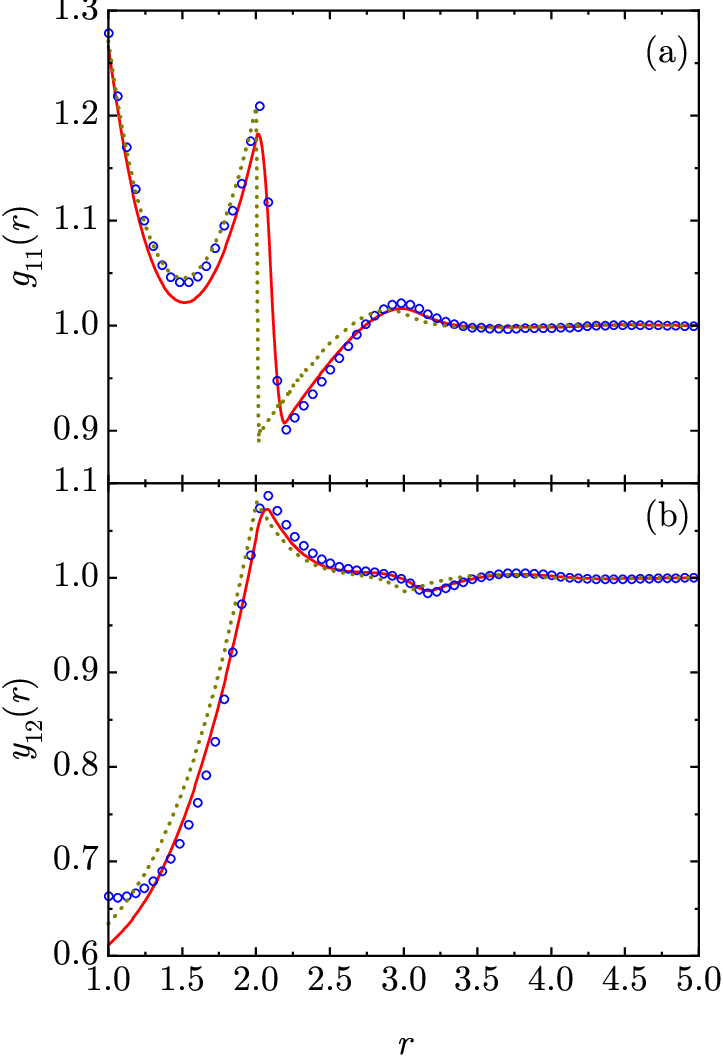}
\captionof{figure}{
Same as in Fig.~\ref{Delta0p1_rho0p1}, except that $\rho=0.4$.
}
\label{Delta0p1_rho0p4}
\end{minipage}

\begin{minipage}[t]{\mpwidth}
\includegraphics[height=\figheight]{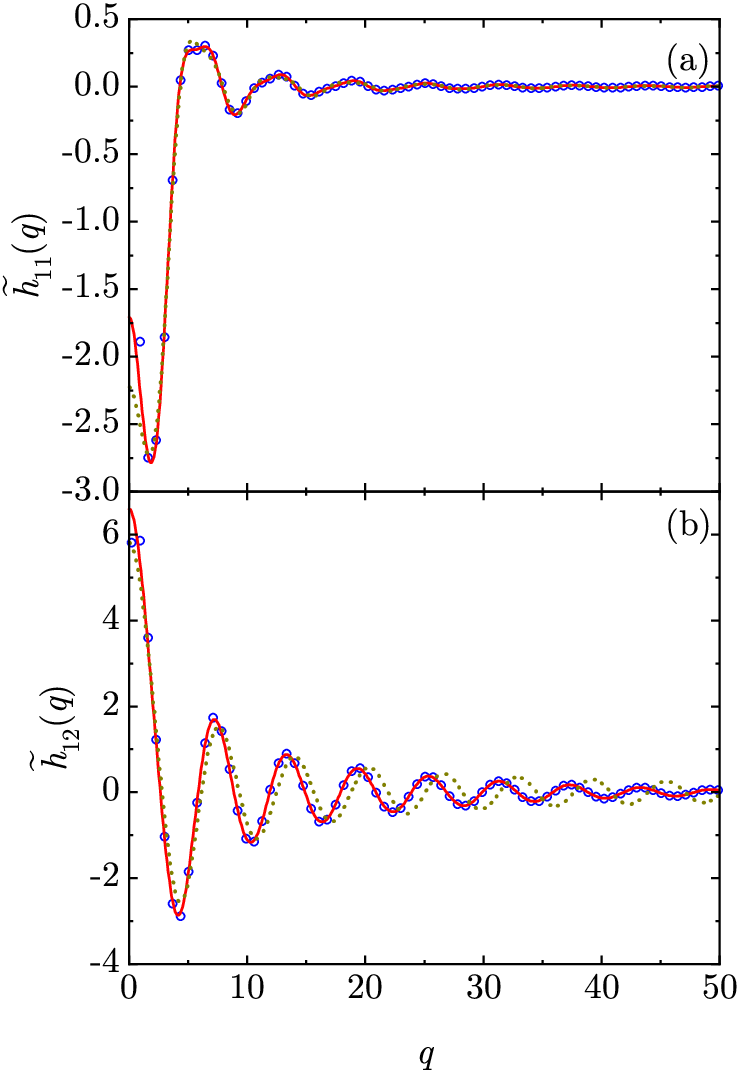}
\captionof{figure}{Plot of (a) $\widetilde{h}_{11}(q)$ and (b) $\widetilde{h}_{12}(q)$ for a binary SW mixture with $x_1=\frac{1}{2}$, $\sigma_1=\sigma_2=1$, $\epsilon_{11}=\epsilon_{22}=0$, $\beta\epsilon_{12}=-2.5$, $\Delta_{12}=0.1$, and  $\rho=0.1$.
The symbols are MC data \cite{Gianmarco26}, the solid lines represent our RFA results, and the dotted lines represent the PY-SHS prediction with the same value of the stickiness parameters:  $\alpha_{11}=\alpha_{22}=0$, $\alpha_{12}=(e^{-\beta\epsilon_{12}}-1)\Delta_{12}=1.118$.
}
\label{htilde_Delta0p1_rho0p1}
\end{minipage}
\hfill
\begin{minipage}[t]{\mpwidth}
\includegraphics[height=\figheight]{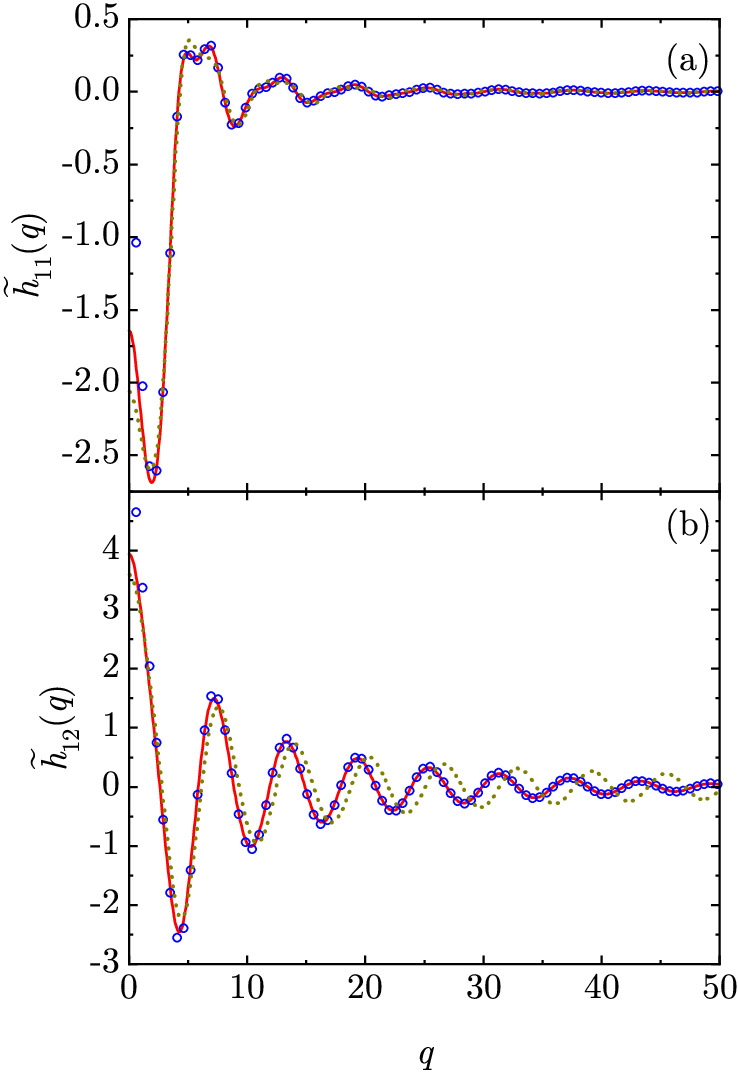}
\captionof{figure}{Same as in Fig.~\ref{htilde_Delta0p1_rho0p1}, except that $\rho=0.2$.}
\label{htilde_Delta0p1_rho0p2}
\end{minipage}

\begin{minipage}[t]{\mpwidth}
\includegraphics[height=\figheight]{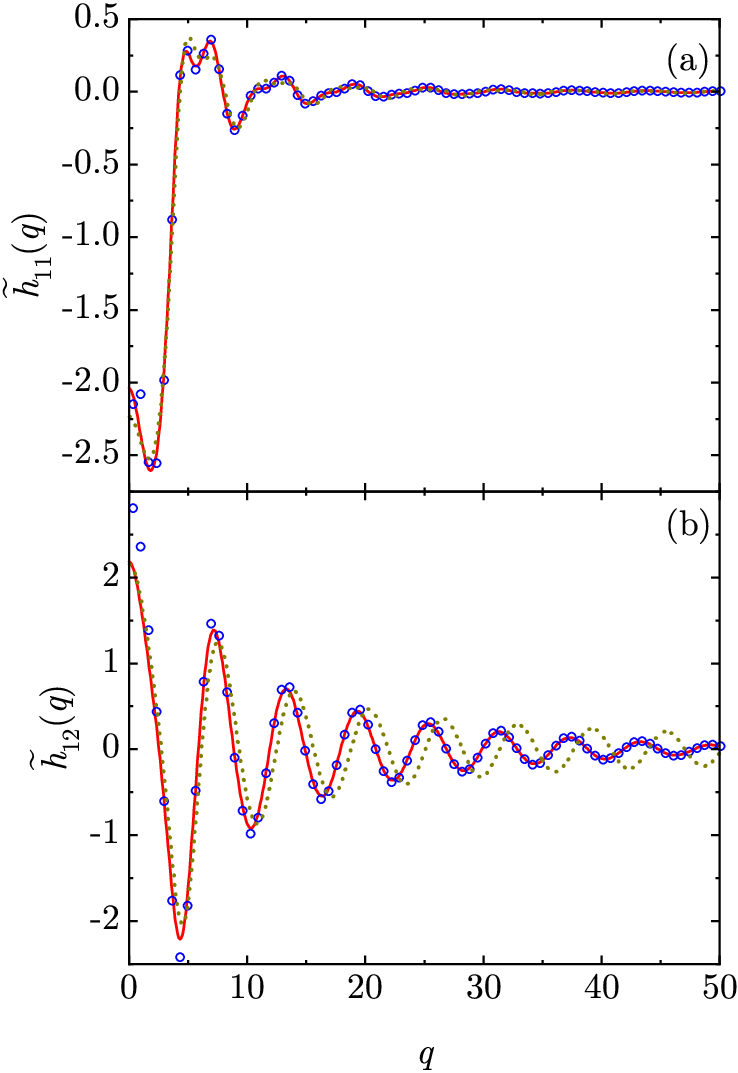}
\captionof{figure}{Same as in Fig.~\ref{htilde_Delta0p1_rho0p1}, except that $\rho=0.3$.
}
\label{htilde_Delta0p1_rho0p3}
\end{minipage}
\hfill
\begin{minipage}[t]{\mpwidth}
\includegraphics[height=\figheight]{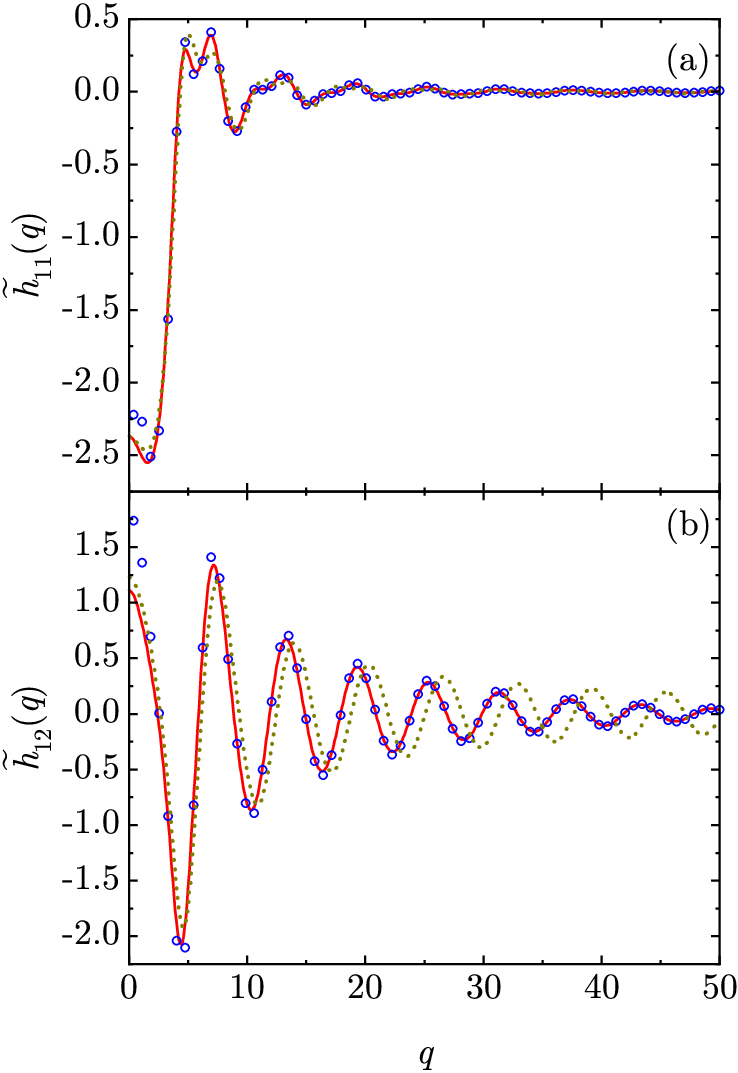}
\captionof{figure}{Same as in Fig.~\ref{htilde_Delta0p1_rho0p1}, except that $\rho=0.4$.}
\label{htilde_Delta0p1_rho0p4}
\end{minipage}

\begin{minipage}[t]{\mpwidth}
\includegraphics[height=\figheight]{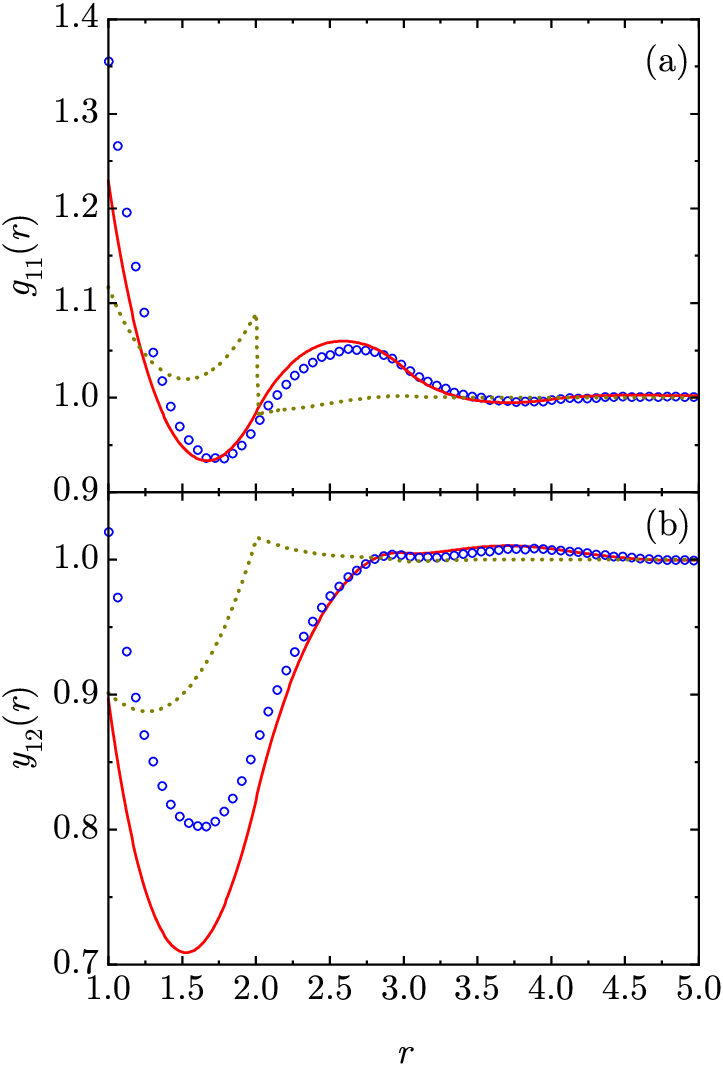}
\captionof{figure}{Plot of (a) $g_{11}(r)$ and (b) $y_{12}(r)$ for a binary SW mixture with $x_1=\frac{1}{2}$, $\sigma_1=\sigma_2=1$, $\epsilon_{11}=\epsilon_{22}=0$, $\beta\epsilon_{12}=-0.5$, $\Delta_{12}=1$, and  $\rho=0.2$. The symbols are MC data \cite{MCMBP23}, the solid lines represent our RFA results, and the dotted lines represent the PY-SHS prediction with the same value of the stickiness parameters:  $\alpha_{11}=\alpha_{22}=0$, $\alpha_{12}=(e^{-\beta\epsilon_{12}}-1)\Delta_{12}=0.649$.
}
\label{fig1aT2_Munao}
\end{minipage}
\hfill
\begin{minipage}[t]{\mpwidth}
\includegraphics[height=\figheight]{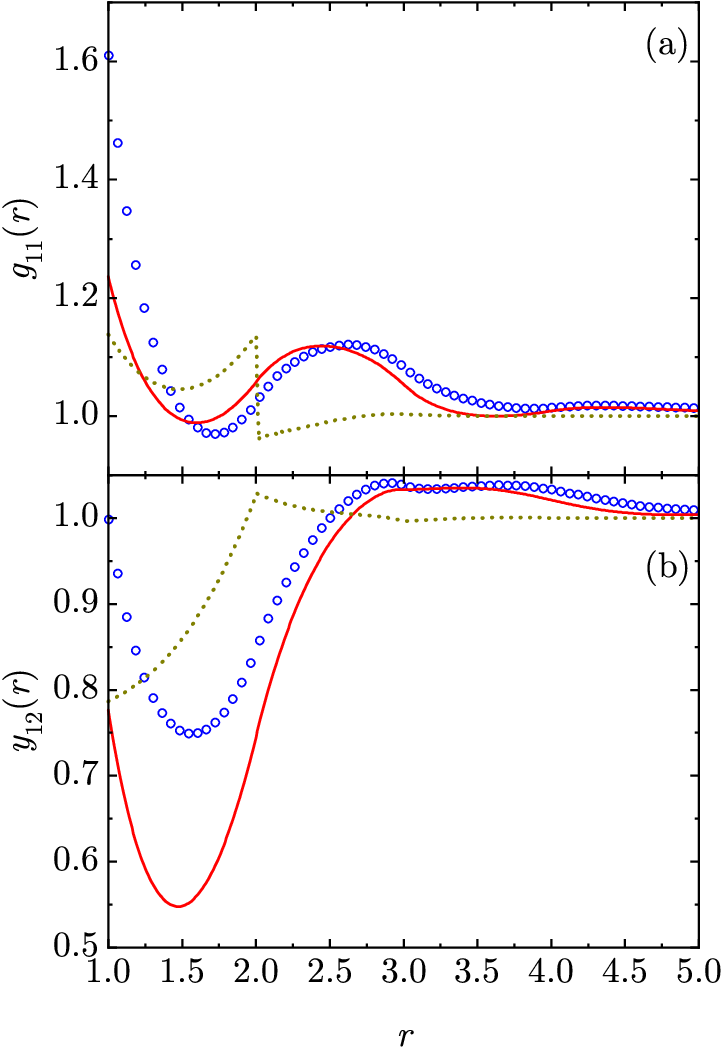}
\captionof{figure}{Same as in Fig.~\ref{fig1aT2_Munao}, except that $\beta\epsilon_{12}=-0.667$ and $\alpha_{12}=(e^{-\beta\epsilon_{12}}-1)\Delta_{12}=0.948$.}
\label{fig1aT1p5_Munao}
\end{minipage}

\begin{minipage}[t]{\mpwidth}
\includegraphics[height=\figheight]{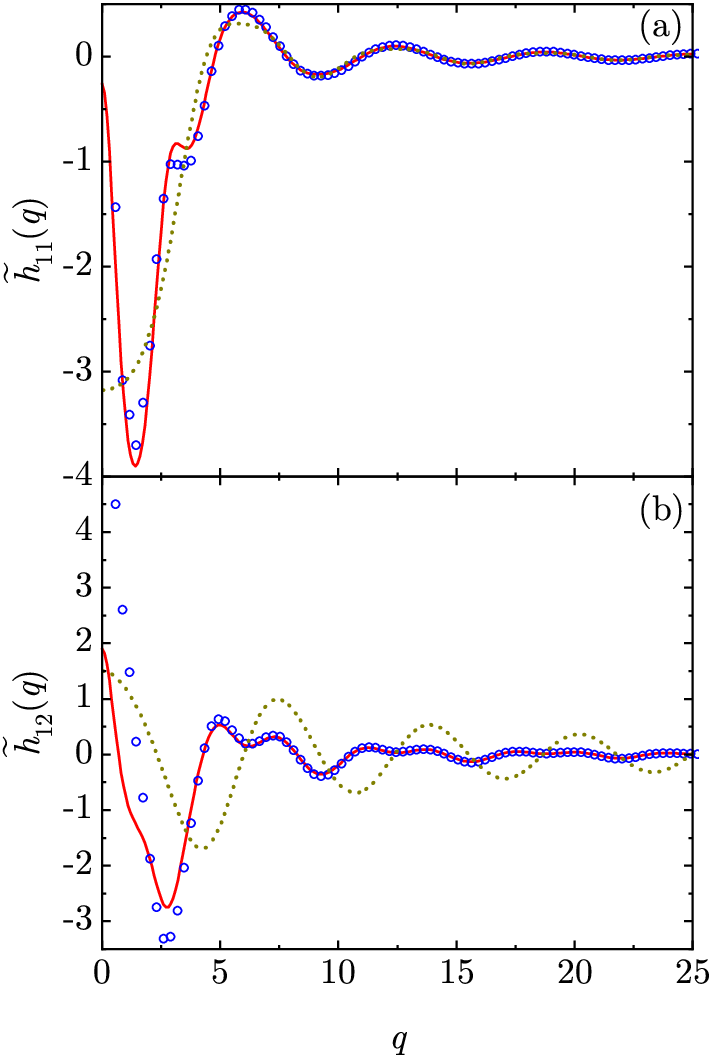}
\captionof{figure}{Plot of (a) $\widetilde{h}_{11}(q)$ and (b) $\widetilde{h}_{12}(q)$ for a binary SW mixture with $x_1=\frac{1}{2}$, $\sigma_1=\sigma_2=1$, $\epsilon_{11}=\epsilon_{22}=0$, $\beta\epsilon_{12}=-0.5$, $\Delta_{12}=1$, and  $\rho=0.2$. The symbols are MC data \cite{MCMBP23}, the solid lines represent our RFA results, and the dotted lines represent the PY-SHS prediction with the same value of the stickiness parameters:  $\alpha_{11}=\alpha_{22}=0$, $\alpha_{12}=(e^{-\beta\epsilon_{12}}-1)\Delta_{12}=0.649$.
}
\label{fig1bT2_Munao}
\end{minipage}
\hfill
\begin{minipage}[t]{\mpwidth}
\includegraphics[height=\figheight]{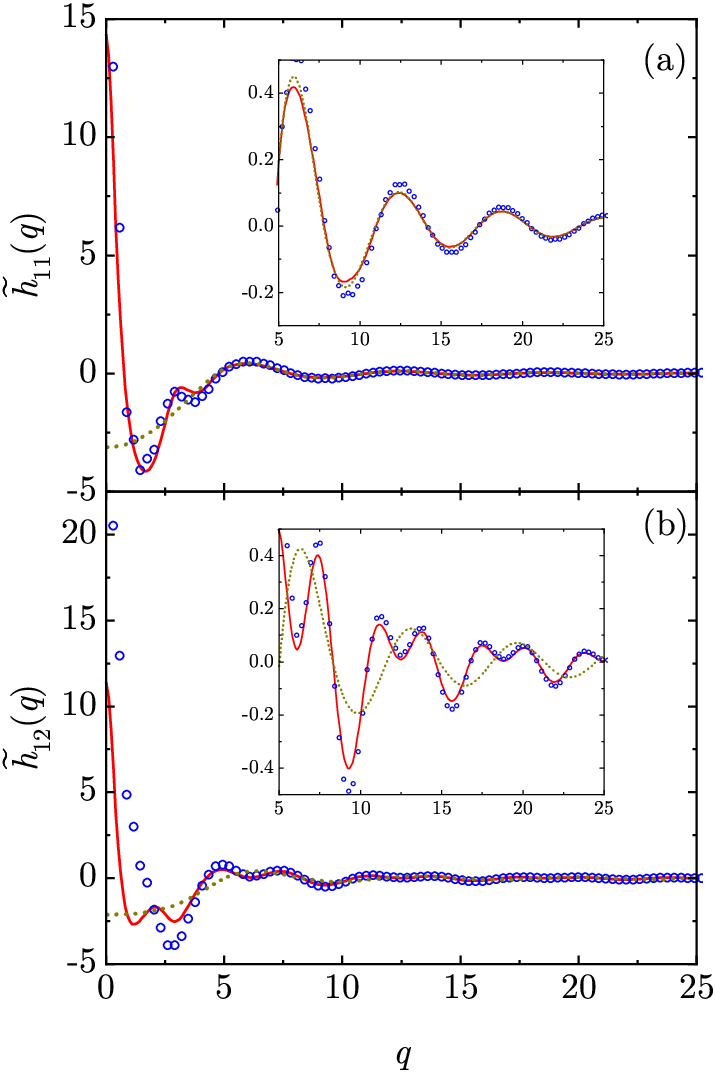}
\captionof{figure}{Same as in Fig.~\ref{fig1bT2_Munao}, except that $\beta\epsilon_{12}=-0.667$ and $\alpha_{12}=(e^{-\beta\epsilon_{12}}-1)\Delta_{12}=0.948$. The insets highlight the region $5\leq q\leq 25$.}
\label{fig1bT1p5_Munao}
\end{minipage}

\begin{minipage}[t]{\mpwidth}
\includegraphics[height=\figheight]{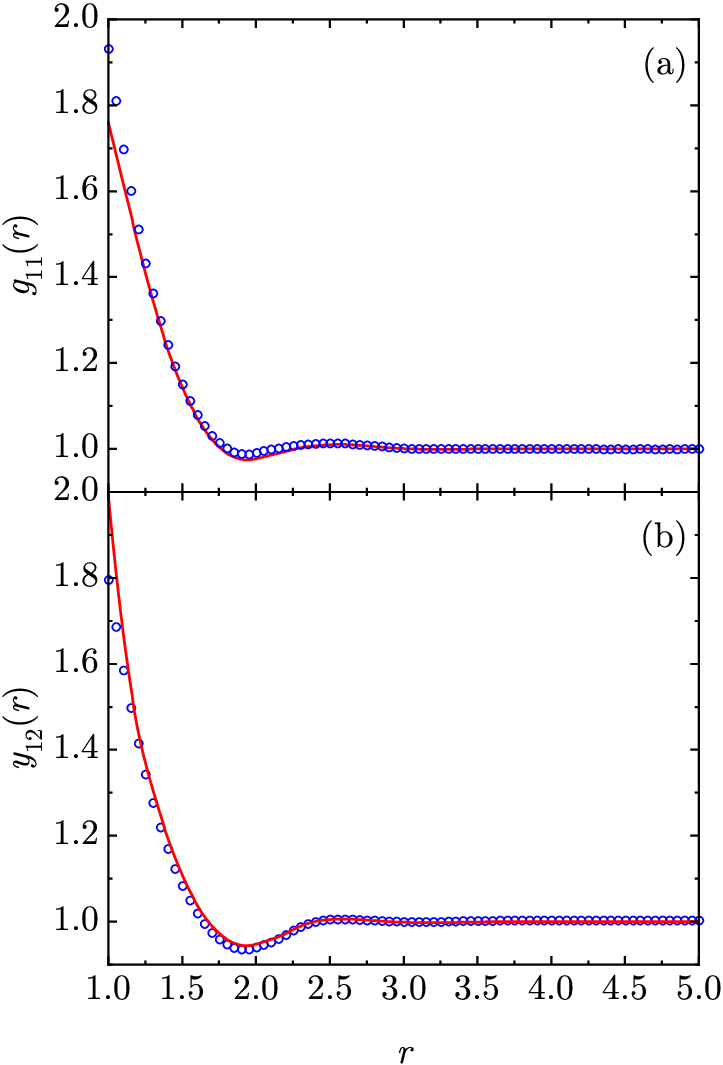}
\captionof{figure}{Plot of (a) $g_{11}(r)$ and (b) $y_{12}(r)$ for a binary SS mixture with $x_1=\frac{1}{2}$, $\sigma_1=\sigma_2=1$, $\epsilon_{11}=\epsilon_{22}=0$, $\beta\epsilon_{12}=1$, $\Delta_{12}=0.25$, and  $\rho=0.3$. The symbols are our own MC data and the solid lines represent our RFA results.}
\label{SS12_Delta0p25_rho0p3}
\end{minipage}
\hfill
\begin{minipage}[t]{\mpwidth}
\includegraphics[height=\figheight]{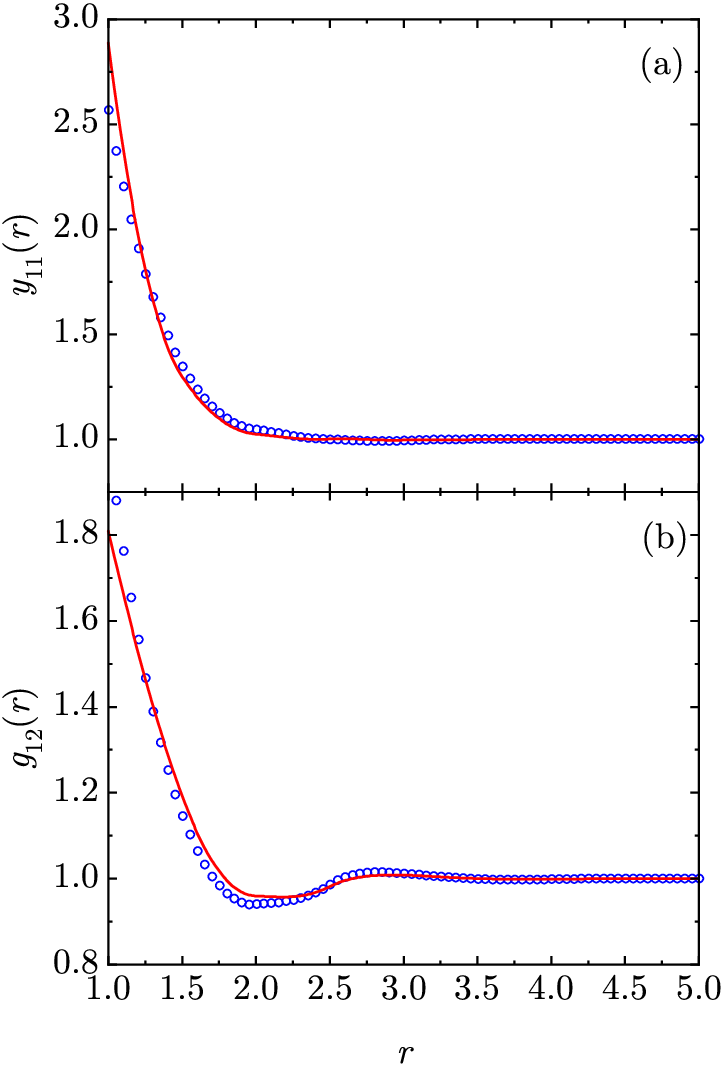}
\captionof{figure}{Plot of (a) $y_{11}(r)$ and (b) $g_{12}(r)$ for a binary SS mixture with $x_1=\frac{1}{2}$, $\sigma_1=\sigma_2=1$, $\beta\epsilon_{11}=\beta\epsilon_{22}=1$, $\epsilon_{12}=0$, $\Delta_{11}=0.5$, and  $\rho=0.3$. The symbols are our own MC data and the solid lines represent our RFA results.}
\label{SS11_Delta0p5_rho0p3}
\end{minipage}

\begin{minipage}[t]{\mpwidth}
\includegraphics[height=\figheight]{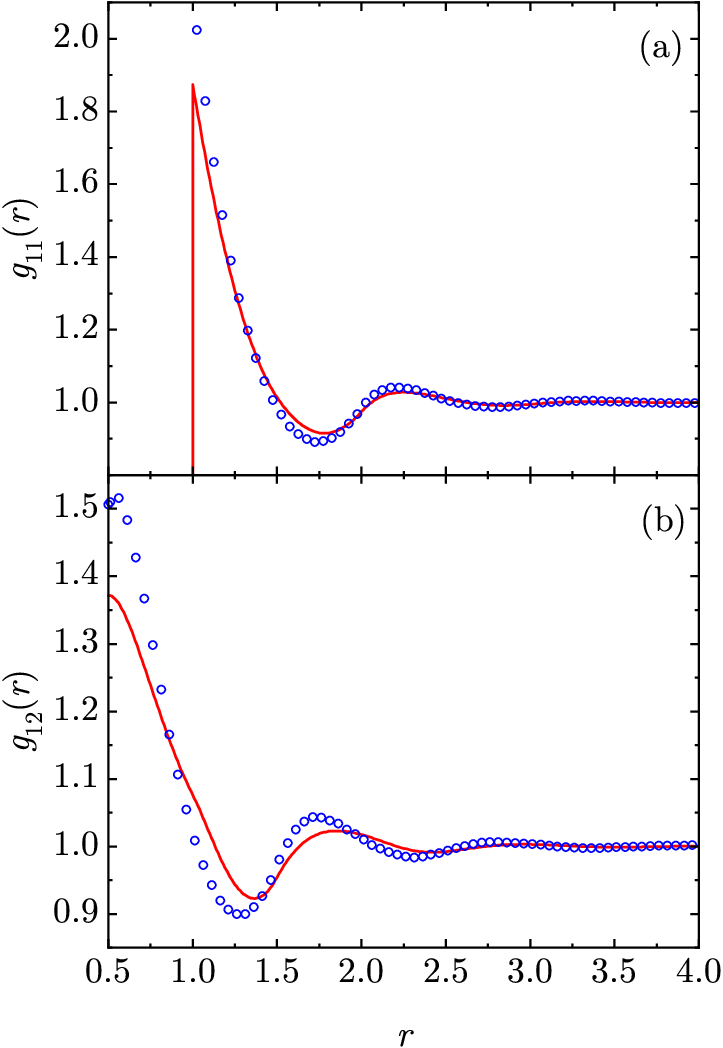}
\captionof{figure}{Plot of (a) $g_{11}(r)$ and (b) $g_{12}(r)$ for a binary NAHS mixture with $x_1=\frac{1}{2}$, $\hs_1=\hs_2=1$, $\hs_{12}=0.5$,   and $\rho=1$. The symbols are MC data \cite{FS11} and the solid lines represent our RFA results obtained with $\beta\epsilon_{11}=\beta\epsilon_{22}=10$ and $\beta\epsilon_{12}=0$.
The value $\beta\epsilon_{1}=\beta\epsilon_{22}=10$ is sufficiently large for the results to be effectively indistinguishable from the NAHS limit; larger values produce no appreciable changes in the results shown.
}
\label{fig6_Riccardo}
\end{minipage}
\hfill
\begin{minipage}[t]{\mpwidth}
\includegraphics[height=\figheight]{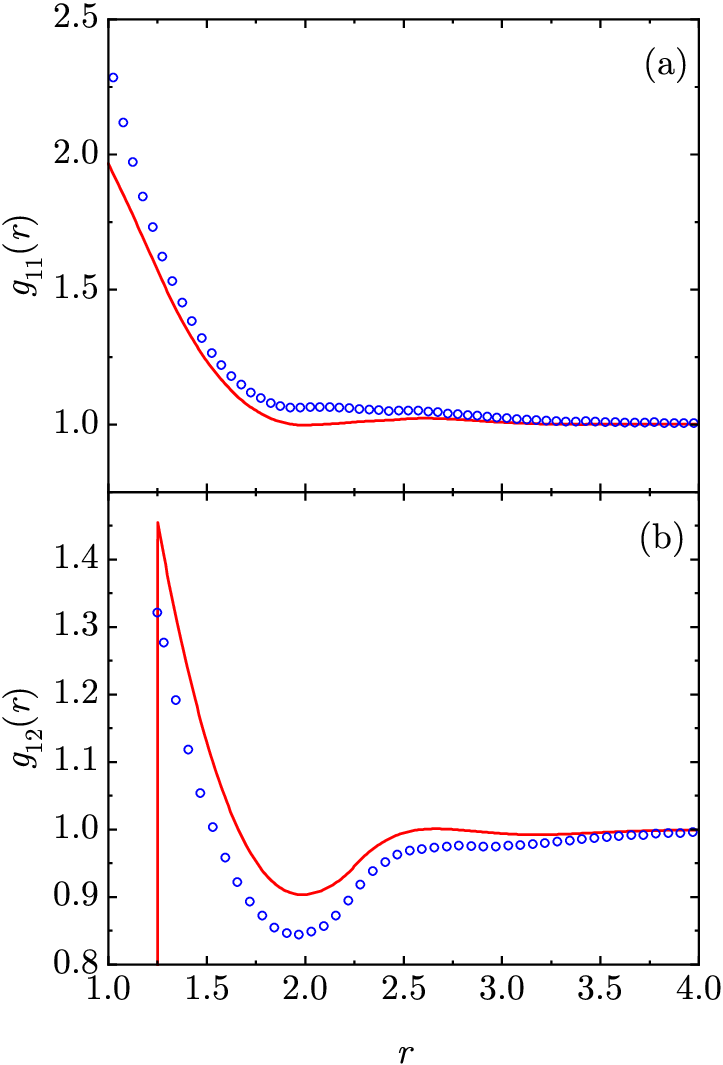}
\captionof{figure}{Same as in Fig.~\ref{fig6_Riccardo}, except that $\hs_{12}=1.25$, $\rho=0.3$, $\beta\epsilon_{11}=\beta\epsilon_{22}=0$, and $\beta\epsilon_{12}=10$.
The value $\beta\epsilon_{12}=10$ is sufficiently large for the results to be effectively indistinguishable from the NAHS limit; larger values produce no appreciable changes in the results shown.}
\label{fig9_Riccardo}
\end{minipage}

\begin{minipage}[t]{\mpwidth}
\includegraphics[height=\largefigheight]{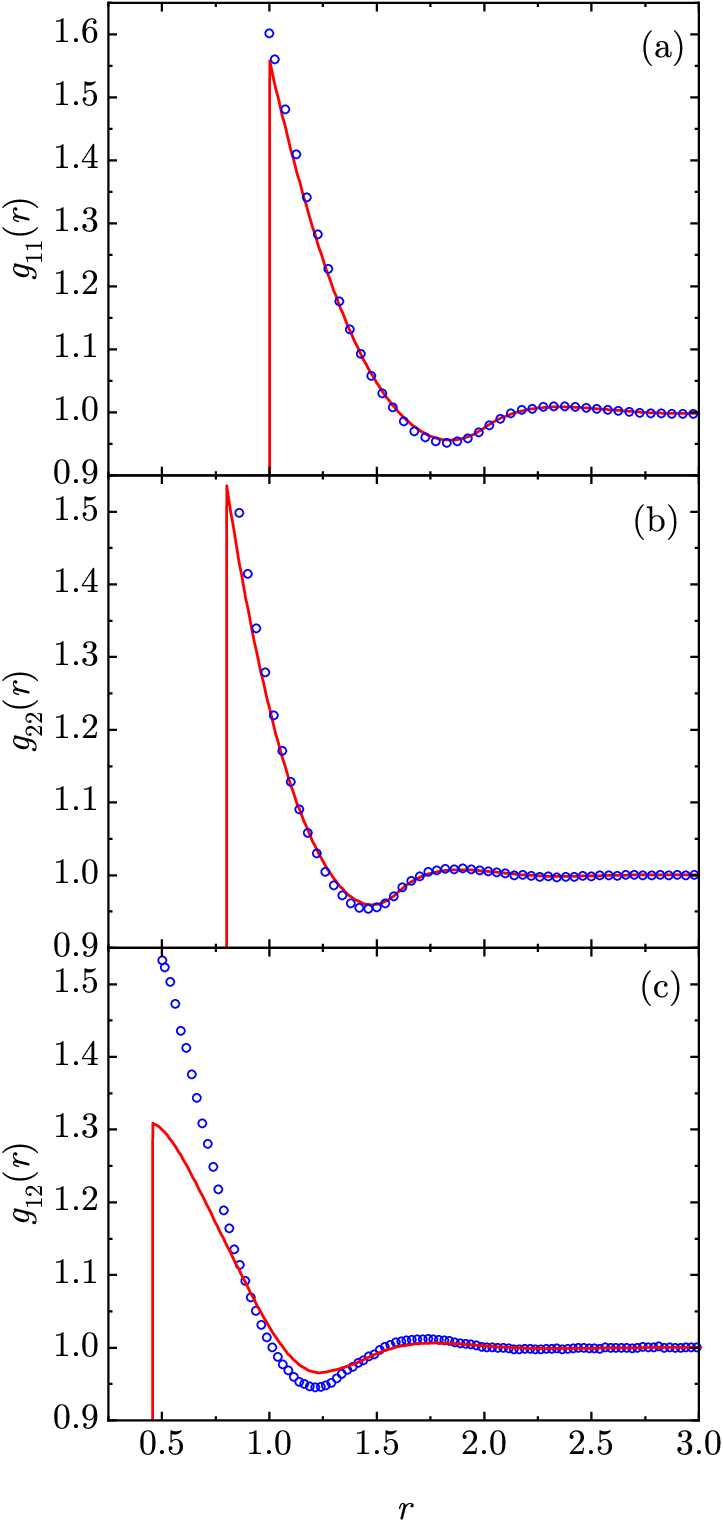}
\captionof{figure}{Plot of (a) $g_{11}(r)$, (b) $g_{22}$, and (c) $g_{12}(r)$ for a binary NAHS mixture with $x_1=\frac{1}{3}$, $\hs_1=1$, $\hs_2=0.8$, $\hs_{12}=0.456$,  and $\rho=1$. The symbols are MC data \cite{FS11} and the solid lines represent our RFA results obtained with $\beta\epsilon_{11}=\beta\epsilon_{22}=10$ and $\beta\epsilon_{12}=0$.
The value $\beta\epsilon_{11}=\beta\epsilon_{22}=10$ is sufficiently large for the results to be effectively indistinguishable from the NAHS limit; larger values produce no appreciable changes in the results shown.
}
\label{fig7_Riccardo}
\end{minipage}
\hfill
\begin{minipage}[t]{\mpwidth}
\includegraphics[height=\largefigheight]{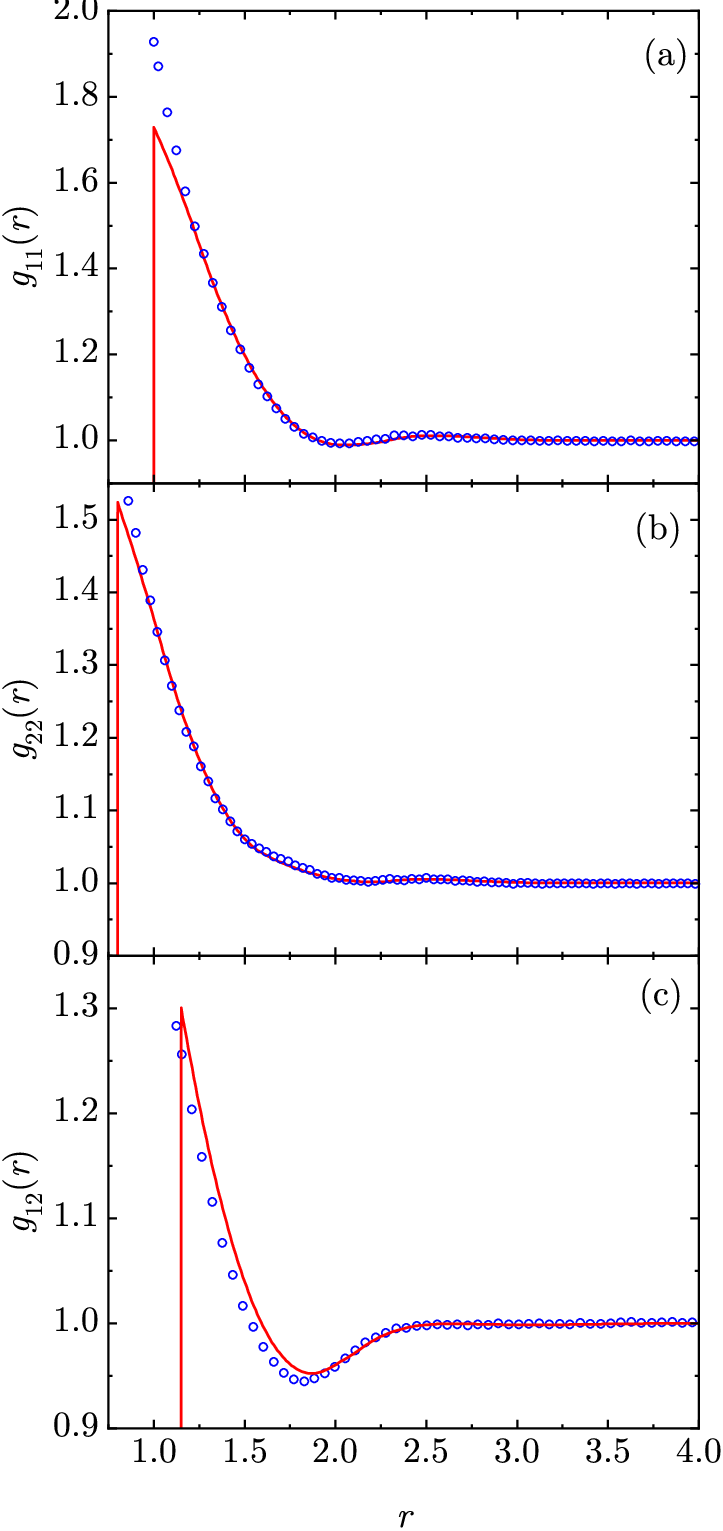}
\captionof{figure}{Same as in Fig.~\ref{fig7_Riccardo}, except that $\hs_{12}=1.15$, $\rho=0.3$, $\beta\epsilon_{11}=\beta\epsilon_{22}=0$, and $\beta\epsilon_{12}=10$.
The value $\beta\epsilon_{12}=10$ is sufficiently large for the results to be effectively indistinguishable from the NAHS limit; larger values produce no appreciable changes in the results shown.}
\label{fig10_Riccardo}
\end{minipage}

\begin{minipage}[t]{\mpwidth}
\includegraphics[height=\largefigheight]{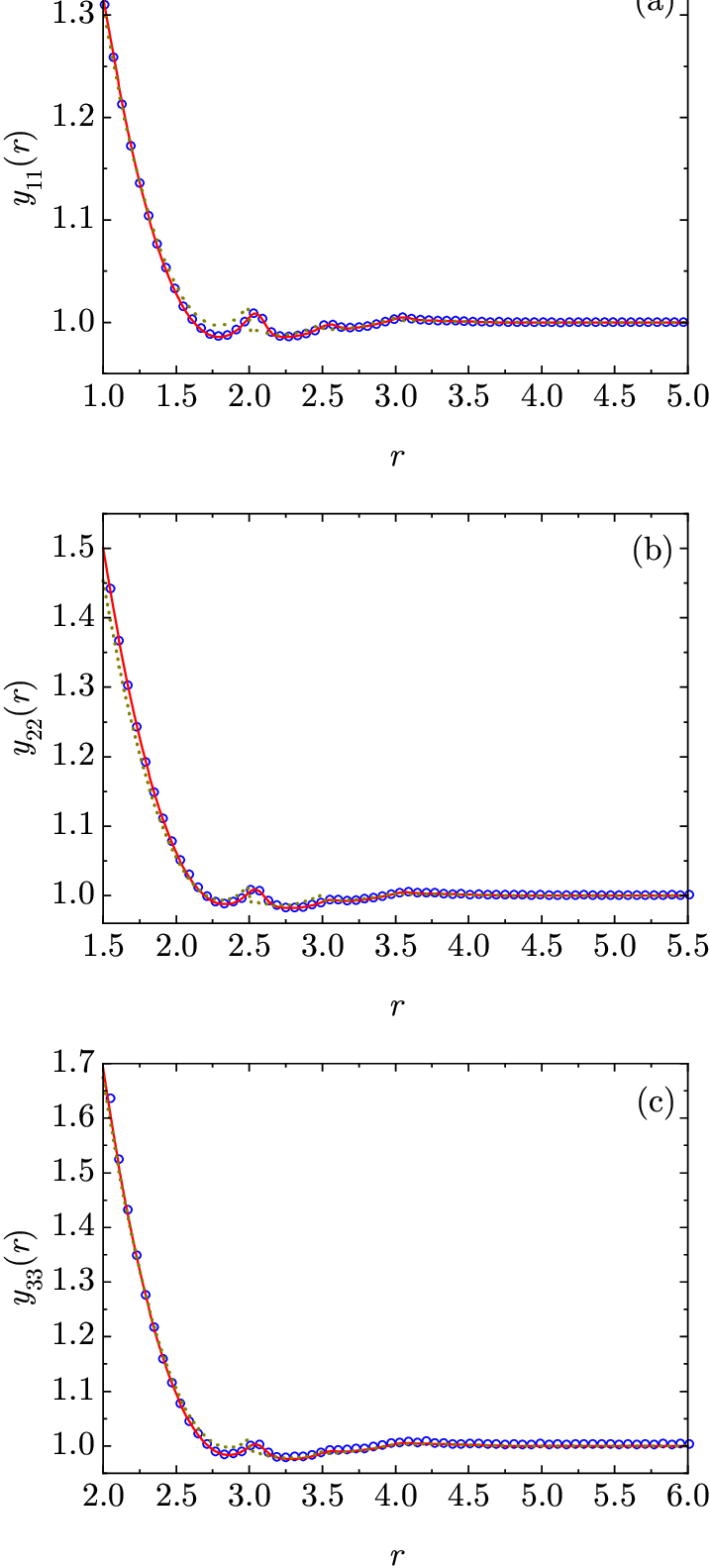}
\end{minipage}
\hfill
\begin{minipage}[t]{\mpwidth}
\includegraphics[height=\largefigheight]{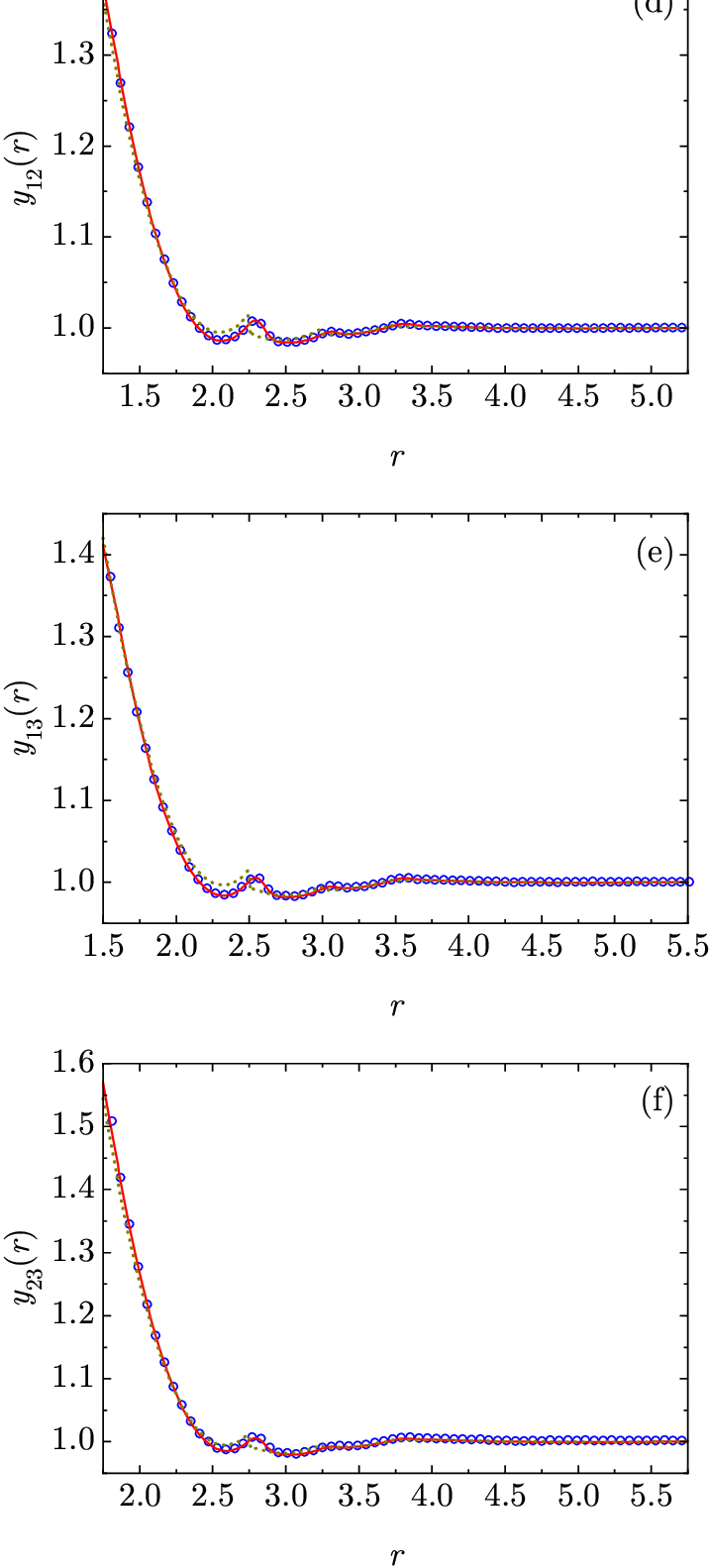}
\end{minipage}
\captionof{figure}{Plot of (a) $y_{11}(r)$, (b) $y_{22}(r)$,  (c) $y_{33}(r)$, (d) $y_{12}(r)$, (e) $y_{13}(r)$,  and (f) $y_{23}(r)$ for a ternary SW mixture with $x_1=0.7$, $x_2=0.2$, $x_3=0.1$, $\sigma_1=1$, $\sigma_2=1.5$, $\sigma_3=2$, $\Delta_{ij}=\Delta=0.1$, $\beta\epsilon_{11}=-1$,  $\beta\epsilon_{12}=-0.9$, $\beta\epsilon_{13}=-0.8$, $\beta\epsilon_{22}=-0.7$, $\beta\epsilon_{23}=-0.6$, $\beta\epsilon_{33}=-0.5$, and $\rho=2$. The symbols are our own MC data, the solid lines represent our RFA results, and the dotted lines represent the PY-SHS prediction with the same value of the stickiness parameters:  $\alpha_{11}=(e^{-\beta\epsilon_{11}}-1)\Delta=0.172$, $\alpha_{12}=(e^{-\beta\epsilon_{12}}-1)\Delta=0.146$, $\alpha_{13}=(e^{-\beta\epsilon_{13}}-1)\Delta=0.123$, $\alpha_{22}=(e^{-\beta\epsilon_{22}}-1)\Delta=0.101$, $\alpha_{23}=(e^{-\beta\epsilon_{23}}-1)\Delta=0.082$, and $\alpha_{33}=(e^{-\beta\epsilon_{33}}-1)\Delta=0.065$.
}
\label{Ternary}

\end{center}

\end{document}